# Generative modeling, design and analysis of spider silk protein sequences for enhanced mechanical properties

Wei Lu[1,2], David L. Kaplan[3], Markus J. Buehler[1,2,4*]

[1] Laboratory for Atomistic and Molecular Mechanics (LAMM), Massachusetts Institute of Technology, 77 Massachusetts Ave., Cambridge, MA 02139, USA

[2] Department of Civil and Environmental Engineering, Massachusetts Institute of Technology, 77 Massachusetts Ave., Cambridge, MA 02139, USA

[3] Tufts University, Medford, MA

[4] Center for Computational Science and Engineering, Schwarzman College of Computing, Massachusetts Institute of Technology, 77 Massachusetts Ave., Cambridge, MA 02139, USA

[*] email: mbuehler@mit.edu

**Abstract:** Spider silks are remarkable materials characterized by superb mechanical properties such as strength, extensibility and lightweightedness. Yet, to date, limited models are available to fully explore sequence-property relationships for analysis and design. Here we propose a custom generative large-language model to enable design of novel spider silk protein sequences to meet complex combinations of target mechanical properties. The model, pretrained on a large set of protein sequences, is fine-tuned on ~1,000 major ampullate spidroin (MaSp) sequences for which associated fiber-level mechanical properties exist, to yield an end-to-end forward and inverse generative strategy. Performance is assessed through: (1), a novelty analysis and protein type classification for generated spidroin sequences through BLAST searches, (2) property evaluation and comparison with similar sequences, (3) comparison of molecular structures, as well as, and (4) a detailed sequence motif analyses. We generate silk sequences with property combinations that do not exist in nature, and develop a deep understanding the mechanistic roles of sequence patterns in achieving overarching key mechanical properties (elastic modulus, strength, toughness, failure strain). The model provides an efficient approach to expand the silkome dataset, facilitating further sequence-structure analyses of silks, and establishes a foundation for synthetic silk design and optimization.

## 1. Introduction

Many biological materials encountered in nature exhibit complex hierarchical structures and unique combinations of material properties. They have hence inspired scientists in material designs and optimization by understanding their underlying features[1–3]. Spider silk is one of the most intriguing biological materials, abundant in nature and this family of protein-based materials has evolved over 300 million years[4–6], and features a complex hierarchical architecture (**Figure 1**), ranging from amino acid chains dictated by spidroin sequences, to protein nanocomposites, to silk fibrils, fibers and webs[1,5,7]. Silk nanofibrils are highly oriented along longitudinal axis but less ordered radially, resulting in mechanical anisotropy along both parallel and normal directions of the fiber axis[8]. The hierarchical nature contributes to the exceptional mechanical properties of spider silks, including toughness, strength, and extensibility, but still lightweight. These properties arise from the presence of stiff crystalline $\beta$-sheets in the nanocomposites, providing silk material stiffness and tensile strength, as well as the disordered extensible semi-amorphous phase and weak hydrogen bonds which enhance extensibility[1,4]. Some spider silks also exhibit supercontraction properties and can shrink up to 50% of its length when exposed to high humidity[9–11] due to the transition from oriented to disorganized silk morphology[12], and this feature is presumed to play a role in tuning mechanical properties during the silk spinning process[13,14]. Furthermore, spider silks are biocompatible and biodegradable[1,15]. Consequently, they have gained interest in various fields, including structural engineering[16,17] and electronics[18,19] due to their mechanical properties, biomedical technologies such as tissue engineering[20,21] and drug delivery[22,23] , and even in art and music[24,25].

In addition, spider silks are multi-functional, with eight main types of silks produced by the corresponding glands, each serving various purposes: (1) Dragline silks form scaffolds that provide structural support for webs; (2) Flagelliform silks are core fibers of capture spirals; (3) Auxiliary spiral silks stabilize and reinforce the overall web structures; (4) Aggregate silks provide sticky aqueous coating for capture spiral; (5) Pyriform silks are used as cement for attachments and joints; (6) Aciniform silks serve the soft inner silks of egg sacs and prey wrapping; (7) Cylindriform silks are designed for constructing egg sacs; (8) Tubuliform silks protect egg sacs by forming an outer coating[26,27]. The spider silk properties are governed by the structure of its main protein building blocks, known as spidroins[28]. Each type of silk is primarily composed of a specific spidroin: (1) MaSp (major ampullate spidroins), (2) Flag (flagelliform spidroin), (3) MiSp (minor ampullate spidroins), (4) AgSp (aggregate spidroin), (5) PySp (pyriform spidroin), (6) AcSp (aciniform spidroin), (7) CySp (cylindrical spidroin), and (8) Tub (tubuliform spidroin)[26–29]. The diversity and evolution of spidroin sequences and their relationship to silk properties has been explored[28,30], as have the terminal domains[31,32], repetitive motifs and regions[33,34], and modifications and paralogs[35,36]. Furthermore, specific functions of different MaSp types have been proposed[28], with MaSp1 contributing to silk strength, MaSp2 enhancing elasticity and supercontraction due to the ratio of amorphous to $\beta$-sheet regions[37], and MaSp3 demonstrating exceptional toughness[28]. Nevertheless, due to the high diversity and complexity of spider silk and spidroins, the sequence-property relationships of spider silks have not yet been fully understood.

Synthetic design for biomaterials and biostructures is appealing to scientists due to their exceptional properties and unique features. Many approaches have been employed to explore this approach, including experimental and theoretical methods[38–40], genetic engineering and motif assembly[41–43], multiscale simulations[44–46], and additive manufacturing[47–49]. Yet, to date, there are limited models available to fully explore sequence-property relationships for analysis and design, especially to achieve novel silk sequences with property combinations that do not exist in nature. In the realm of spider web structures, synthetic webs and web-inspired designs has been explored using deep learning[5,15,50], topology optimization[51,52], additive manufacturing and simulations[44,53,54].

In this paper we solve this challenge with a custom deep generative model based on a transformer architecture. Deep learning techniques have advanced in prediction and generation tasks across various data types, encompassing natural language data, graphs, and images[55,56]. Several deep generative models have been developed for material and mechanical problems[57], including the design of three-dimensional nature-inspired materials[58], including graphene[59], proteins[60,61], other chemicals[62], and the development of a large language model (LLM) for multiple forward and inverse mechanics problems[63]. As for spider web-related studies, web properties have been predicted using graph neural networks and GraphPerceiver transformer attention models[50,64,65], and web designs can be generated through diffusion model and autoregressive transformer with different neighborhood representations[5].

Research related to silk classification and prediction tools have been implemented to assess properties and design patterns in silk fabrics[66,67], as well as the conformation of silk proteins[68]. However, there is limited exploration of generative techniques within the domain of spider silks and their associated protein sequences using deep learning methodologies. Such strategies would more directly allow the creation of novel experimental designs and protocols for future research and design targets. Solving this challenge is the objective of this paper.

Numerous deep learning algorithms[56] have been developed for sequence generation, such as Recurrent Neural Networks (RNNs), Convolutional Neural Networks (CNNs), Variational Autoencoders (VAEs), and Generative Adversarial Networks (GANs)[69]. More recent techniques have emerged with impressive generation capacity. For instance, the diffusion model[62,70] introduces Gaussian noise in diffusion process and iteratively reconstructs the output in denoising process[70]; and transformers with attention mechanisms, capture intricate relationships within sequential data or tokens[71]. Notable examples include the encoder model BERT[72] (Bidirectional Encoder Representations from Transformers) and the autoregressive model GPT[73–75] (Generative Pre-trained Transformer). These models can be pre-trained through Masked Language Models and Left-to-Right Language Models respectively, followed by fine-tuning to enhance model performance. Moreover, each model has variations tailored for specific generation purposes and incorporating unique model features, such as ALBERT[76], ProtBert and ProtBert-BFD; GPT-NeoX[74] and GPT-J[77].

Furthermore, the development of the Silkome data collection[28], which compiles spidroin sequences and material properties, serves as a valuable source for training models of spider silk proteins. Silkome collects spidroin data from transcriptome assemblies of 1,098 species, with a total of more than 11,000 unique sequences, and includes detailed material testing data of dragline silk properties from 446 species[28] (and a total of around 1,000 sequence-property associations). In this work, we take additional steps to refine and prepare the dataset for training, as detailed in **section 4**. Significant protein motifs that influence the measured silk properties were also studied previously[28], which we utilize to validate the generation capability of our model, as discussed in **section 3**.

### ***Motivation for this study***

Synthetic silk design can have a significant impact regarding material improvement and innovations, biotechnology design, sustainability and cost reduction for economic opportunities, on multiple research areas and industries, including material and structural engineering[39,78,79], tissue engineering[80,81], biomedical applications related to wound healing[82,83] and drug delivery[84,85], as well as synthetic meat and other agriculture applications. However, challenges exist in the design of synthetic spider silk domains since there are limited data on spidroin sequences, due to high diversity and complexity of spider silks and spidroins, which constrains the exploration of the intricate relationships between the sequence structures and the resulting functions of spidroin[28]. Moreover, although there have been studies exploring synthetic spider silk design through experimental approaches[42,86] and genetic engineering and motif assembly[22,41,42]. The high costs for synthetic designs to assess properties post synthesis, purification and processing prompt the need for advanced computational modeling methods to be more fully utilized for spider silk protein sequence analysis. Incorporating advanced modeling tools would improve research efficiency compared to current sequencing methods[28,87], as well as enhance design reliability and expand design opportunities. We propose a generative modeling framework that employs an autoregressive transformer to design spider silk protein sequences and to predict their associated fiber-level mechanical properties. The key contributions of this work are outlined below:

- A dataset comprising 1,033 protein sequences is constructed, sourcing from the Silkome[28] dataset, to develop associations of protein sequences with various properties, including protein-level information (e.g., family, genus, species, sex, etc.) and fiber-level mechanical properties (e.g., toughness, elastic modulus, etc.). For the purpose of this study, we focus on the protein sequences and their related mechanical properties, but the dataset can be used for a variety of downstream applications.
- A sequence-based generative modeling method is developed, leveraging advanced deep learning language modeling techniques based on attention mechanisms, to enhance the efficiency of synthetic spider silk protein design. The model is evaluated for self-consistency, and is proficient in generating novel and reliable spider silk protein sequences. The model provides a cost-effective means to expand the spidroin dataset, facilitating further study of sequence-property relationships of spider silks.
- We find that the method attains reasonable predictive accuracy for mechanical properties of spider silks. The framework hence lays a robust foundation for various downstream tasks including *de novo* spidroin generation, synthetic silk design, and silk property optimization. Implementation of the model allows design space exploration for spider silk materials, synthetic silk design with lower impact on environments, and property optimization for silk-based products.
- The model can be expanded to predict other multilevel properties related to spider silk and proteins, and can be adapted for other different protein or biomaterials featuring similar hierarchical structures, to target various design purposes across multiple domains.

The paper is organized as follows. We first introduce spider silk and spidroin, as well as the motivation of this work (this section). We then analyze the generation and property prediction results from generative modeling, including a thorough assessment of the prediction performance and the self-consistency of the model. The third section summarizes the work, key insights including mechanisms, and concludes the impact and outlines potential future tasks. The last section discusses the methodology implemented, including the dataset and model construction, as well as the modeling and training details.

## 2. Results and Analysis

In this work, we explore the relationship between the protein sequences of spider silks and their corresponding fiber-level properties, such as toughness and strength. Our model is able to generate novel spidroin sequences and predict their associated properties. The 8-dimensional mechanical property sets for generation inputs are defined in "**Dataset Construction**" in **Section 4, Materials and Methods**. To assess the generative capability of the generative model, we analyze the performance through three varied mechanical property sets employed as target input conditions. In addition, five property sets are selected to generate novel spidroin sequences, to evaluate the model's self-consistency through novelty check, classification of generated protein sequences, protein property assessment, comparison of folding configurations, and motif analysis. The exact target property values for these selected sets listed in **Table 1.**

The three initial property sets for prediction performance analysis are randomly selected to encompass a range of property values. The five property sets for self-consistency analysis are selected following specific selection criteria, to encompass common, rare, and non-existent mechanical behavior of spider silks, to rigorously assess the adaptability and reliability of our method. The selection process are detailed as follows. As indicated in red in **Figure 2**, we focus on elastic modulus and toughness value pairs for the selection: Set 1 with common elastic modulus and toughness pair (0.2 & 0.25), set 2 and 3 with existing but rare value pairs with relatively high E (0.75 & 0.55) and high toughness (0.45 & 1.00) respectively, set 4 with non-existent pair and with elevated property values of elastic modulus and toughness (0.8 & 0.9). Set 5 instead centered around common strength and toughness value pairs (0.27 & 0.24). For the remaining property values, the medians are employed.

***Prediction Performance***

**Figure 3** provides a comparison between the predicted property values (in red) of the generated novel sequences and the corresponding ground targeted properties (in blue). The bar plots represent the results obtained from the generation with three distinct input property sets (details in **Table 1**). For each generation, the results with the highest R2 values are selected. The three example tests demonstrate a reasonable model prediction performance, with R2 values of 0.8764, 0.8369, and 0.6889, respectively. The output R2 values, obtained from three randomly selected input property sets, demonstrate reasonable correlations between the predicted and target property values, signifying the potential utility of the model for designing spider silk sequences with desired properties. These results further show the flexibility of the model to adapt to various property requirements, suggesting that it can be utilized for a wide range of spider silk designs with different mechanical property constraints.

**Figure 4** presents a similar plot comparing the predicted and ground targeted properties for the five selected target property values (property details in **Table 1**). Spidroin sequences are generated and selected (sequences in **Table S1**) with best output R2 values of 0.8899, 0.5640, 0.7167, 0.7843, and 0.7751 for property set 1 to 5 respectively. Reasonable prediction performance is indicated, with an average R2 of 0.75, and for property sets featuring non-existent values (set 4), a comparable level of performance can be achieved (R2=0.7843), which implies that the model can effectively handle scenarios where the target mechanical properties have unique value combinations, and show the reliability of the model for spidroin sequence design with optimal or tailored properties. The comparison for the five selected property sets, with not only common property value combinations but also rare and non-existent ones, demonstrates the robustness and versatility of the model for property prediction performance. Moreover, it is noted that the selected property sets that with common value pairs (set 1 and set 5) tend to yield higher prediction performance, but the property sets containing rare value pairs (set 2 and set 3) exhibit a slightly lower correlation, which suggests that the prediction capability of the model may be more adapt within the common value property combinations, and this understanding can guide the selection of property sets for specific design goals.

***Novelty Assessment and Protein Classification***

While we exclusively selected novel sequences during the generation process when comparing them with the MaSp dataset, it remains essential to assess their novelty and protein classification within the broader set of protein sequence databases. The five novel spider silk sequences generated based on the five target property sets are now analyzed through BLAST searches. A total of 10 similar, existing spider silk sequences are selected based on the analysis, with two for each of the 5 generated sequences (designated as BLAST 1 and BLAST 2 here). All protein sequences are summarized in **Table S1**, with the sources of ten existing sequences indicated. The selection criteria for similar sequences involve relatively high query cover and percent identity, as well as comparable sequence length when contrasted with the generated novel sequences.

To evaluate novelty, the highest values and the common value ranges for both query cover (QC) and percent identity (id%) are summarized and analyzed in **Table S2**, where QC serves to assess the alignment of sequences, and id% signifies the degree of composition similarity. Sequences exhibiting similarity values below 50% to 60% are considered novel according to established criteria[61,88]. The generated sequences are assessed to be novel within the broader protein dataset, with detailed discussions for all five sequences summarized in **Table S2**, which suggests the capability of model in designing sequences with unique properties that either do not exist or have not been observed in nature. Moreover, through sequence comparison for five property sets, most similar existing sequences from BLAST searches are categorized as MaSp, which aligns with the targeted spidroin type. This result indicates that the generative modeling methods are capable of capturing the key characteristics of specific types for spider silk proteins.

***Protein Property Analysis***

Protein properties for five novel spidroin and ten similar searches are compared in **Table S3** and **Figure 5**. **Table S3** contains the sequence length and other properties, including molecular weight (MW), instability index (II)[89], and Isoelectric point (pI), calculated using ProtParam[90], here through an implemented Python tool package[91]. The comparison for these properties for each property set is visualized in **Figure 5(a)**, the amino acid percentage and the secondary structure fraction are compared in **Figure 5(b) and 5(c),** respectively. The secondary structure fraction refers to the ratio of amino acid tend to be in helix (V, I, Y, F, W, L), turn (N, P, G, S) and sheet (E, M, A, L)[91]. In general, a consistent trend is observed across these property values for the generated sequences and the similar existing ones within each property set. Though value variations exist, the choice of existing sequences for comparison from the BLAST searches can potentially influence the nature of the comparison. The property consistency indicates the generated sequences are biologically meaningful, which supports the validity of the generated sequences regarding structural and biological features. Additionally, the consistent alignment implies that the model can effectively capture key characteristics and properties of spider silk proteins, thus reinforces the reliability of the modeling approach for novel sequence generation, and its applicability in biomaterial design.

***Molecular Structure Comparison***

To compare the folding configurations, we employed AlphaFold2[92] (AF2) to fold the five generated spidroin sequences and one of their related protein search (BLAST 1). The folded configurations are provided in **Figure 6**. We note that these folded configurations are only meant as a comparative means to assess relative changes across distinct sequences. This is because AF2 will likely not predict correct in-situ structures as seen in a silk fiber; and reflecting this hypothesis, the overall folding performance for spider silk proteins using AF2 yields relatively low pLDDT[93,94] (per-residue local distance difference test) scores, ranging from 40 to 60 (values detailed in **Table S3**). This might due to the limited availability of spidroin protein sequences in the training data of AF2.

While noting these limitations we use the structures to explore some molecular insights. We observe similarities comparing the configurations of novel sequences and the BLAST 1 sequences. For example, for property sets 2 to 4, resemblances are observed regarding the alignment of alpha-helices and the overall structure and coils. On the other hand, for property set 1, the overall shapes exhibited less similarity, while the presence of beta-sheets was evident in the folding of both sequences. Thus, the accuracy of AF2 predictions for spider silk folding configurations may not be optimal, the presence of visual similarities between generated and existing sequences implies that our generation model is capable of capturing certain structural features for spider silk protein, and can generate non-existent but valid spidroin sequences. Furthermore, employing the generation method to compare the structural differences can potentially uncover common structural motifs or variations that may influence the silk's mechanical properties. This can enhance the understanding of the structure-property relationship of spider silks, and aid in the design of spider silks with tailored properties.

***Motif Analysis towards Mechanistic Understanding***

Spidroin motifs with evident impact on key mechanical properties are selected based on the insights derived from **Table 1** in the Silkome paper[28]. Here we provide a summary of analyzed motifs in **Table S4**, and assume that the specific classification of MaSp type (e.g., MaSp1) spidroin has minimal impact on the motif analysis. **Figure 7** depicts the motif analysis for all five property sets, from top to bottom respectively. **Figure 7a** compares the motif counts for generated sequences and their two similar searches (BLAST 1 & 2). A recurring trend of similarity is evident between novel sequences and existing ones across most property sets, particular in sets 1, 3, and 4, which feature common property values and optimized non-existent values. The generated sequences share common structural motifs with natural spider silk sequences, which verifies their structural validity and biological relevance. However, the congruence is diminished in set 2 for some motifs, while we acknowledge that the choice of BLAST search sequences could influence this comparison.

Furthermore, the consistency of motif impact is investigated and we find that sequences with higher toughness (sets 2, 3, and 4) display a decreased count of the "T_neg_1 & 2" motif; whereas, sequences with elevated E values, particularly in set 4, exhibit an increased count of the "E_pos_2" motif. Additionally, the KDE plots in **Figure 7b** and **7c** illustrate the relative positioning pattern of specific motifs, for two motif examples "SB_pos_2" and "E_pos_2", respectively. Similar sequence patterns are observed across most property sets. Especially, sequences with higher E values (sets 2, 4, and also 3) demonstrate a higher level of consistency, which might be attributed to the inherent elasticity that MaSp possesses compared to other types of spider silks. These observations validate the impact of motifs in mechanical characteristics of spider silk, as suggested in[28]; as well as demonstrates that the generative method can effectively produce sequences that align with motif impacts, encompassing both composition and positioning, thus contributing to functional properties similar to those observed in natural spider silks.

In summary, the motif analysis provides additional validation the reliability of the generated spidroin sequences and affirms the self-consistency and authenticity of the generative modeling method, thereby providing robust evidence of the method's efficacy.

## 3. Conclusion and Outlook

In this work we report a pretrained and fine-tuned model, the Silk Proteome Generative Pretrained Transformer. We used the MaSp dataset that features 1,033 sequences along with corresponding fiber-level properties, sourced from the Silkome dataset[28]. A spidroin sequence generative modeling method was developed and used to design novel spider silk protein sequences to achieve target mechanical properties. We were able to generate novel silk sequences with property combinations that do not exist in nature. Through a careful analysis of associations of sequences and associated properties, we were able to develop a deep understanding the mechanistic roles of sequence patterns in achieving overarching key mechanical properties (elastic modulus, strength, toughness, failure strain).

Our generative model yields promising predictions, with an R2 value reaches 0.85 when comparing target properties with the properties predicted for the designed sequences. Moreover, the generated sequences are novel and upon further analysis through BLAST searches are shown to feature common design principles shared with similar protein classes. We provided a detailed analysis of query cover and percent identity to measure sequence alignment and composition similarity[61,88]. In addition, we have shown that our model provides excellent consistency with known features, as evidenced by comparing the properties of the novel sequences with similar existing, *i.e.*, natural, ones. Overall, highly consistent trends are observed across both generated and existing sequences regarding protein properties, including the instability index, the amino acid percentage, and the fraction of secondary structures. Moreover, structural similarity of overall folded configurations for novel and associated similar natural sequences is confirmed. The most salient motifs, which have been evaluated for their influence on the mechanical behavior of spider silk[28], show consistent impacts across the analyzed sequences, suggesting that our generative model has accurately learned critical construction principles of silk sequences for certain mechanical properties at the fiber level.

There are many impacts of this work. For instance, the model developed here can effectively expand the spidroin dataset, facilitating sequence-structure analysis of spider silks, to spawn future experimental work. The framework established here serves a robust foundation for downstream tasks including *de novo* spidroin generation, synthetic silk design, and silk property optimization. Specifically, it allows the expansion of design space for silk-based materials, enabling the exploration of novel silk materials with improved functional properties or intriguing geometries, such as lightweight structural components, biomedical materials, or aerospace applications. The synthetic silk design approach has the potential to reduce environmental impact associated with traditional sericulture and silk product manufacturing processes. The property optimization could lead high performance fabric products, such as sports clothing with higher durability and extensibility.

Beyond these impacts, the construction of our model as an end-to-end tool can serve as an example for other complex biological and non-biological materials. The pretrained silk proteome model can be fine-tuned easily for other properties across the various hierarchical levels of silk structure, including solubility for drug design, biodegradability for sustainable consumer products, and biocompatibility for silk-based medical devices. Furthermore, the spidroin generative model can also be extended to other proteins (e.g., collagens, keratins, enzymes), and biomaterials (e.g., silkworm silk, bone, hair) with similar hierarchical structures, targeting various design objectives across multiple domains. While the focus here was primarily on mechanical properties, a wider range of functional material outcomes can be considered in future iterations, such as material failure modes, optical performance, electronic features, and bio-interactions in general related to biocompatibility and degradability, among other features, can be realized.

## 4. Materials and Methods

### *Dataset construction*

In this study, we focus on modeling the MaSp type of spidroin, which serves as the primary components of silk that form spider web scaffolds and contributes to high mechanical properties of spider silks, as discussed in **section 1**. The dataset we employ is constructed and curated based on the *Silkome* dataset[28]. This dataset comprises a total of 1,033 data, with each protein-level sequence paired four corresponding fiber-level mechanical properties: toughness, elastic modulus (E), tensile strength, strain at break. Additionally, we incorporate the standard deviation (SD) values of these properties, derived from measurements. The inclusion of SD values in our modeling helps to capture the inherent variability in these mechanical properties, and to assess the relative consistency and reliability of both measured and predicted property values.

All eight data values are all normalized to be between 0…1 (normalization enhances the training with faster convergence, and simplifies the value comparison for data analysis). For each property, the minimum number and maximum values are initially identified across the entire dataset (not limited to MaSp sequences). Subsequently, normalization is applied by subtracting the minimum value from each data point and then dividing them by the calculated range. These scaling parameters are saved to ensure consistent data utilization so that any normalized value can easily be converted to original units). **Table S5** summarizes all normalization scaling parameters for the 8 properties.

The resulting 8-dimensional properties set, used as a conditioning feature for model fine-tuning and as input for conditional generation, is described as follows:

$$[\text{toughness}, \text{SD of toughness}, \text{E}, \text{SD of E}, \text{strength}, \text{SD of strength}, \text{strain}, \text{SD of strain}]$$

As for the MaSp dataset, the distributions of its normalized mechanical properties and standard deviations are shown in the right and the left panel in **Figure 2,** respectively, along with the correlations. In general, uniform distributions are observed among the properties, with toughness and strength exhibiting a stronger positive correlation compared to the others. Furthermore, **Figure S1** illustrates the correlations between each mechanical property and its respective standard deviation. The toughness and elastic modulus are more positively correlated with their standard deviation values, than strength and strain. It suggests that the variability or uncertainty in measurements tends to increase as the property values increase, particularly for toughness and elastic modulus, which may be attributed to certain factors affecting the spider silk properties or the measurement process as the property values increase.

***Model architecture: Silk Proteome Generative Pretrained Transformer (SilkomeGPT)***

A custom Byte-Pair Encoding (PBE) tokenizer with up to 50,000 tokens (minimum frequency=2) is trained based on a set of ~800,000 protein sequences of up to 1,024 length including the silkome dataset and other known proteins (collected from the AlphaFold2 prediction database from a  variety of organisms including Human, M. jannaschii, Mouse, Maiz, and many others (https://alphafold.ebi.ac.uk/, constructed from sequences for UP000000805, UP000008816, UP000001584, UP000000625, UP000002485, UP000001450,UP000000559, UP000002311, UP000008153, UP000002195, UP000000803, UP000002296, UP000001940, UP000005640, UP000002494, UP000000589, UP000000437, UP000006548, UP000008827, UP000007305, UP000059680, and  Swiss-Prot). The development of a custom protein sequence tokenizer the ability of the model to capture patterns effectively. The tokenizer learns the vocabulary and performs tokenization as input to the transformer model (the PBE tokenizer thereby learns the statistical patterns in the silk sequence data and merges the most frequently occurring byte pairs into new tokens).

We use a Generative Pretrained Transformer (GPT) architecture featuring a decoder-only self-attention transformer model, with rotary positional embedding, inspired by the GPT-NeoX architecture (**Figure 1b-c**). The model features 12 layers, 8 attention heads, a hidden dimension of 1024, and an intermediate size of 4096 in the feed-forward layer (the model has 253.6 million parameters). The model is pretrained and fine-tuned from scratch as outlined in the next section.

***Training process and other hyperparameters***

The generation model is first pretrained using protein sequences sourced from a substantial pretraining set of ~600,000 protein sequences (a subset of the sequences used for tokenizer training with lengths of up to 600 amino acid residues) that include the silkome sequences. By including both general protein sequences and silk sequences, the model learns diverse structural features across species, materials and organisms (see list above). Pretraining is achieved via next-token predictions (we also explored denoising pretraining strategies albeit no significant performance enhancement was found). The pretraining task is realized via this input, as an example:

```
Sequence<GGDSGAAAAAAAAADGGRGGYGGLGRGGDSGAAAAAAAAADGGRGGYGGLGSGGDSGAAAAAAAGEDNEGIRYGPGGSSGAAA
AAAAAGEGDSGLGYGPRSESGAASAAAAAAASDGLGGSYGPEFVRSLSSNLMSSEHFLSTFSGSITQNRALSASLALARNYALQSGLNSA
AANAMLNLVNRYISEVGSFAEAESYANALARAIAEGLGNAATRNNFGTLGSTGGRRGTGYMRRIGSTSAAASASDAGVNGPGYGAENGL
GSGAAAAAAAAGEGNEGLGYGPGGASGAAAAAAAAADGGRGGLYGRRGGYGGDSGAAAAAAAAGESNDGLGYGSGGASGAAAAAAAAADDGRG
GYGGVGSGLGEESGAAAAAAAAAGEGDEGFGYGPGGASGAAAAAAAAADGGRRRLYGRRGGFGGDSGAAAAAAAAAGEGNEGLGYGPGGASE
AAAAAAAADDGSVGFGGIGRGLGGDSDAAAAAAAAAGEGNEGLGYGPGGSSGAAAAAASADGGRGGLYGRRRGNGGDSGAAAAAAAAGEGG
LGSGGAPSITFIRSVTPSVSRVNSFTSSLVSGGSLNVGALPEIISAGINEIASYSSGLSECEISGQVLLDIIASLLHILRYSNIGSVDY
STVGETNGYLSSVLGGY>
```

We use a batch size of 48 and 4 gradient accumulation steps, with a linear learning rate schedule and a warmup phase.

We then fine-tune the model against the generative and silk property prediction tasks, for sequences up to 768 amino residues length. Fine-tuning features three tasks: "Sequence" (same as during pretraining), "CalculateSilkContent" and "GenerateSilkContent". The sequence task here only includes silk sequences.

In the training data, we use this bidirectional association of properties designated via the task tokens, to yield, as an example, for the forward task:

```
CalculateSilkContent
<AAAGGAGQGGYGGQGAGQGAAAAAAGGAGQGGYGGQGAGQGAGAAAAAAGGAGQGGYGGLGSGQGGYGGQGAGAAAAAAAAGGAGQGG
YGGLGSGQGGYGGQGAGAAAAAAGGAGQGGYGGLGGQGAGQGSGAAAAAAGGAGQGGYGGQGAGQGAGAAAAAAGGAGQGGYGGLGGQG
AGQGAAAAAAGGAGQGGYGGQGAGQGAGAAAAAAGGAGQGGYGGLGSGQGGYGGQGAGAAAAAAGGAGQGGYGGLGGQGAGAAAAAAGG
AGQGGYGGQGAGQGAAAAAAGGAGQGGYGGQGAGQGGYGGQGAGAAAAAAGGAGQGGYGGLGGQGAGQGAGAAAAAAGGAGQGGYGGQG
AGQGAGAAAAAAGGAGQGGYGGLGGQGAGAAAAAAGGAGQGGYGGQGAGQGGYGGQGSGAAAAAAAAGGAGQGGYGGLGSQGAGQGAGA
AAAAAGGAGQGGYGGQGAGQGAGAAAAAAGGAGQGGYGGQGAGQGAGAAAAAAGGAGQGGYGGQGAGQGAGAAAAAAGGAGQGGYGGLG
SGQGGYGGQGAGAAAAAAGGAGQGGYGGQGAGAAAASAAASRLSSPEASSRVSSAVSNLVSSGPTNSAALSNTISSVVSQISASNPGLS
GCDVLVQALLEVVSALIHILGSSSIGPVNYGSASQSTQIVGQSVYQALG>
[0.327,0.356,0.261,0.287,0.437,0.190,0.220,0.301]
```

And as an example for the generative inverse task:

```
GenerateSilkContent<0.327,0.356,0.261,0.287,0.437,0.190,0.220,0.301>
[AAAGGAGQGGYGGQGAGQGAAAAAAGGAGQGGYGGQGAGQGAGAAAAAAGGAGQGGYGGLGSGQGGYGGQGAGAAAAAAAAGGAGQGG
YGGLGSGQGGYGGQGAGAAAAAAGGAGQGGYGGLGGQGAGQGSGAAAAAAGGAGQGGYGGQGAGQGAGAAAAAAGGAGQGGYGGLGGQG
AGQGAAAAAAGGAGQGGYGGQGAGQGAGAAAAAAGGAGQGGYGGLGSGQGGYGGQGAGAAAAAAGGAGQGGYGGLGGQGAGAAAAAAGG
AGQGGYGGQGAGQGAAAAAAGGAGQGGYGGQGAGQGGYGGQGAGAAAAAAGGAGQGGYGGLGGQGAGQGAGAAAAAAGGAGQGGYGGQG
AGQGAGAAAAAAGGAGQGGYGGLGGQGAGAAAAAAGGAGQGGYGGQGAGQGGYGGQGSGAAAAAAAAGGAGQGGYGGLGSQGAGQGAGA
AAAAAGGAGQGGYGGQGAGQGAGAAAAAAGGAGQGGYGGQGAGQGAGAAAAAAGGAGQGGYGGQGAGQGAGAAAAAAGGAGQGGYGGLG
SGQGGYGGQGAGAAAAAAGGAGQGGYGGQGAGAAAASAAASRLSSPEASSRVSSAVSNLVSSGPTNSAALSNTISSVVSQISASNPGLS
GCDVLVQALLEVVSALIHILGSSSIGPVNYGSASQSTQIVGQSVYQALG]
```

The fine-tuning training set includes all known pairs of sequence and properties.

All code is developed in PyTorch, within the Hugging Face ecosystem, for further details see[95]. All machine learning training is performed using an Adam optimizer[96], with a learning rate of 0.0002. Data is available via the original Silkome paper or via our GitHub page: https://github.com/lamm-mit/SilkomeGPT. The pretrained and fine-tuned models can be utilized in future work to solve other tasks; and the flexibility of the language-based modeling strategy allows for straightforward addition of other modeling tasks (e.g. solubility, other properties, genus/family classification tasks, and others).

### Conflict of Interest

The authors declare no conflict of interest.

### Authors' Contributions

M.J.B. developed the overall concept and the algorithm, designed the generative model, the initial codes, oversaw the work. W.L. contributed codes and analysis, prepared the dataset, analyzed the results, and drafted the paper. W.L., D.L.K., M.J.B analyzed the data and edited and wrote the manuscript. All authors participated in discussions about the research ideas, results and conclusions.

### Acknowledgements

We acknowledge support from the MIT Generative AI Initiative, NIH and USDA. Related support from the MIT-IBM Watson AI Lab, MIT Quest, and Google Cloud Computing, is acknowledged. We acknowledge Keiji Numata for the development of the Silkome dataset.

**Data Availability**

The data that support the findings of this study are available from the corresponding author upon reasonable request. Codes and data are available at: https://github.com/lamm-mit/SilkProteomeGPT.

**Supplementary Information**

Supplementary figures and tables referred in this paper are provided as **Supplementary Information**.

## Figures

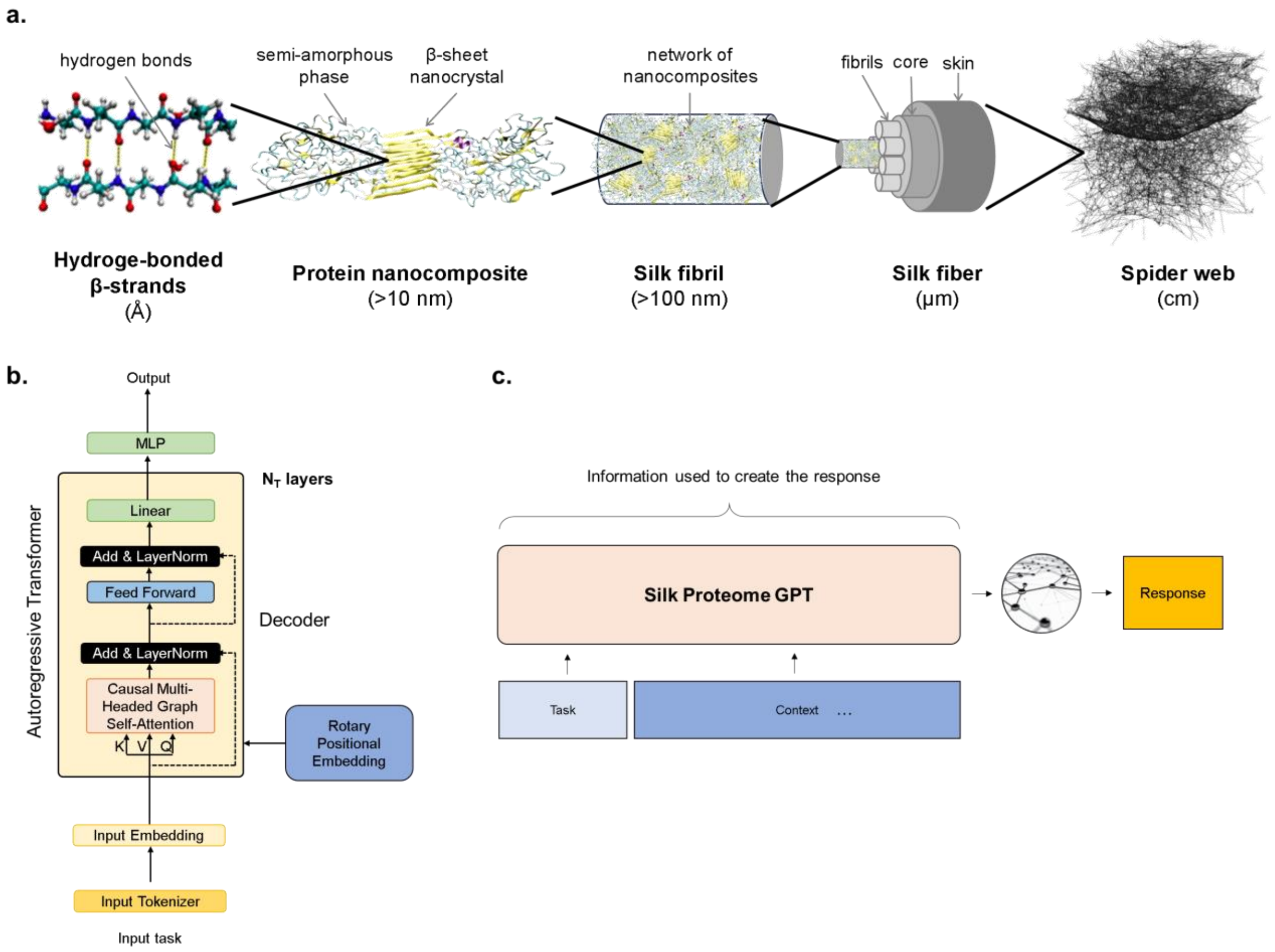

**Figure 1: Hierarchical structure of silk and the SilkomeGPT modeling approach used, based on a transformer architecture**. Panel a: Schematic of hierarchical structure of spider silk, from nanoscale hydron bonded chains to macroscale spider web structures (adapted from[7] with permission). Hydrogen-bonded β-strands assemble into β-sheet nanocrystals, which are embedded in semi-amorphous phases. These nanocrystals unite to compose protein composites, constituting the building blocks of silk fibrils. Bundles of the silk fibrils form the spider fibers, then the macroscale spider webs structures. The protein-level structures are represented by amino acid sequences, and the fiber-level properties are measured in[28] and utilized in this work. Panel b: Summary of the autoregressive decoder-only transformer model architecture, with rotary positional embedding. Panel **c**: Overview of modeling approach. A query including the task and relevant context is used to create the responses, with interactions among all elements considered via graph-forming attention mechanisms. Tasks included in this work include the petraining "sequence" task, as well as a "calculate" and a "generate" task. The "calculate" task predicts a set of mechanical properties based on a given sequence, and the "generate" task yields a silk sequence with associated mechanical properties (details see **Materials and Methods**).

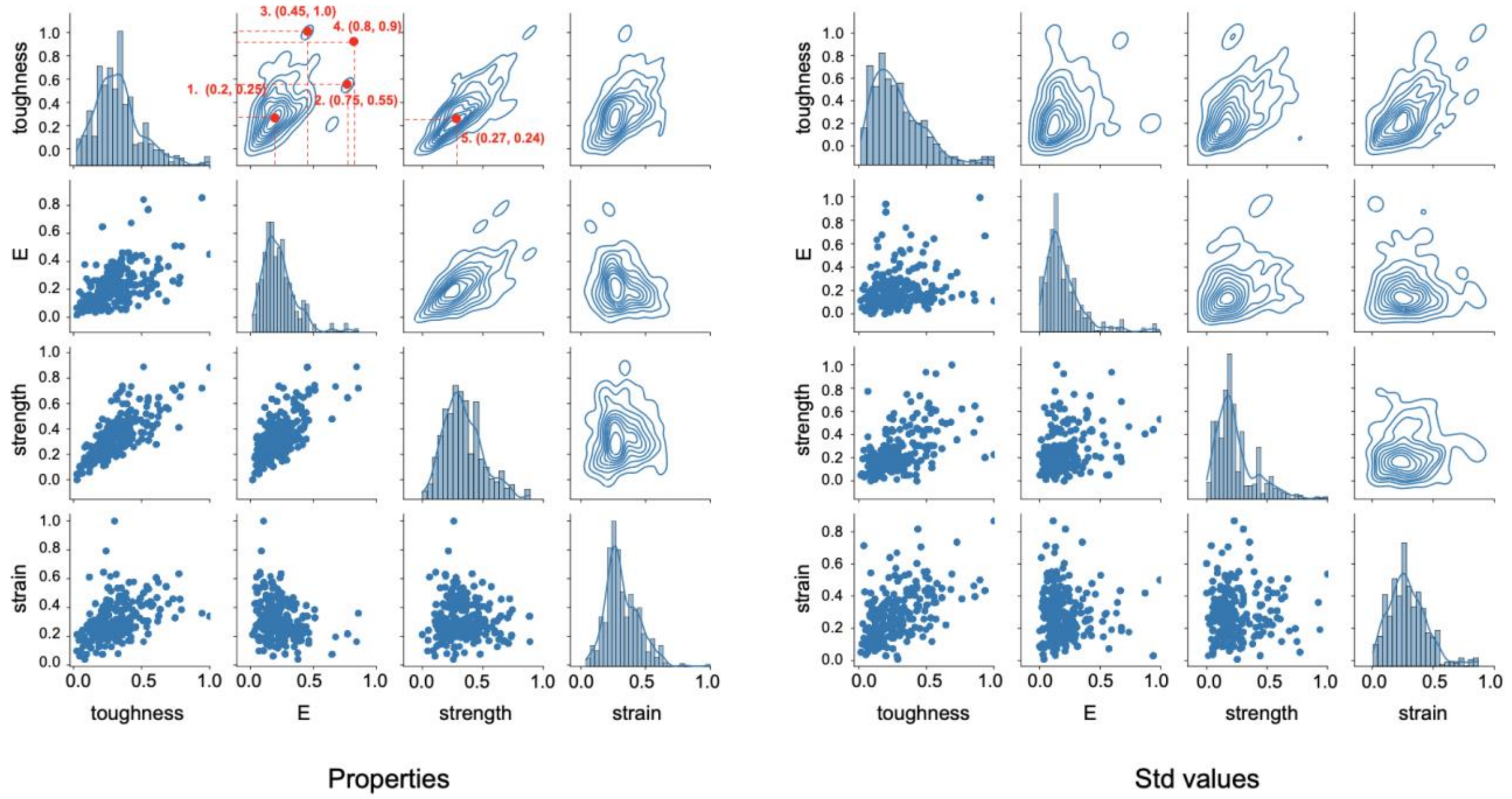

**Figure 2. Analysis of the MaSp spider silk property data, based on the data originally reported in**[28]**.** We depict property distributions as histograms with KDE plots (diagonal), and property correlation in scatter plots (lower region) and KDE plots (upper region), of normalized mechanical properties (right panel) and their standard deviations (left panel). In general, uniform distributions are observed, and toughness and strength are relatively positively correlated. Five target property sets are selected (as marked in red) to generate novel sequences for further comparison and analysis. We focus on elastic modulus and toughness value pairs: set 1 with common elastic modulus and toughness pair (0.2 & 0.25), set 2 and 3 with existing but rare value pairs with relatively high modulus *E* (0.75 & 0.55) and high toughness (0.45 & 1.00) respectively, set 4 with non-existent pair and with elevated property values for elastic modulus and toughness (0.8 & 0.9). Set 5 centered around common strength and toughness value pairs (0.27 & 0.24). For other remaining property values, the medians are employed. The specific property values for each sets are as follows: set 1: [0.250, 0.252, 0.200, 0.151, 0.310, 0.203, 0.293, 0.265], set 2: [0.550, 0.252, 0.750, 0.151, 0.310, 0.203, 0.293, 0.265], set 3: [1.000, 0.252, 0.450, 0.151, 0.200, 0.203, 0.293, 0.265], set 4: [0.900, 0.252, 0.800, 0.151, 0.310, 0.203, 0.293, 0.265], set 5: [0.240, 0.252, 0.206, 0.151, 0.270, 0.203, 0.293, 0.265]. The 8-dimensional property set are defined in **Dataset Construction** in **Section 4**, and specific values of the property sets are listed in **Table 1**.

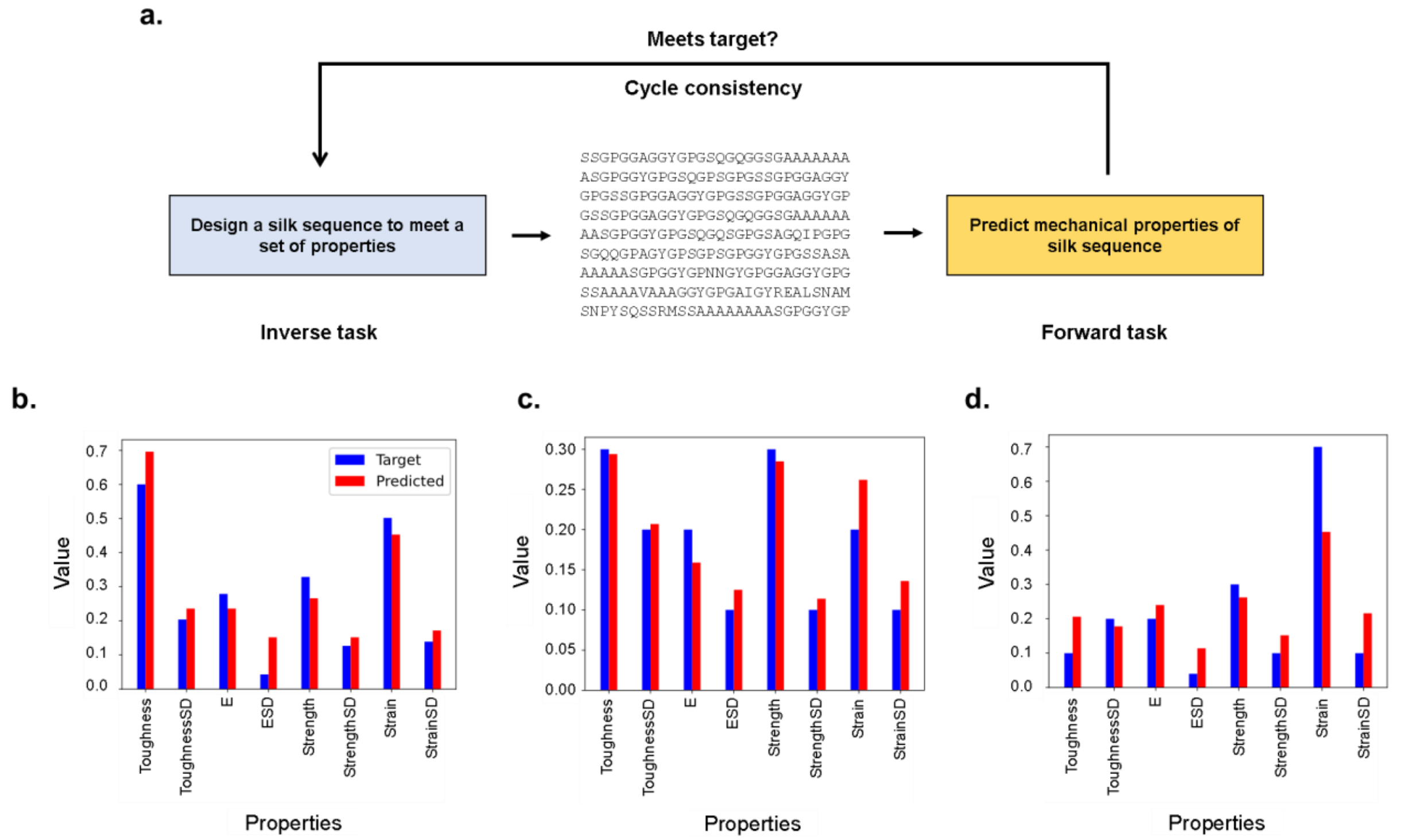

**Figure 3. Comparison of predicted properties and ground truth for the three property sets, achieved using a cycle-consistent generative design approach.** Panel **a** shows a schematic flowchart of how the inverse and forward tasks are combined, and panels **b-d** show bar plots provide a comparison between the predicted property values (in red) of the generated novel sequences and the corresponding target properties (in blue). The bar plots are from generation with three different input targeted property sets, with [toughness, SD of toughness, E, SD of E, strength, SD of strength, strain, SD of strain] equal to [0.600, 0.204, 0.279, 0.043, 0.329, 0.127, 0.502, 0.139], [0.300, 0.20, 0.2, 0.1, 0.3, 0.1, 0.20, 0.1], [0.100, 0.20, 0.2, 0.04, 0.3, 0.1, 0.70, 0.1], respectively. For each generation, the results with the highest R2 values are selected from a set of sampling attempts. The three example tests demonstrate good performance, as indicated by the R2 values: 0.8764, 0.8369, and 0.6889, respectively. The output R2 values, obtained from three randomly selected input property sets, demonstrate moderate correlations between the predicted and target property values, signifying the potential utility of the model for designing spider silk sequences with desired properties. The results also illustrate the flexibility of the model to adapt to various property requirements, suggesting that it can be utilized for a wider range of spider silk designs with different property constraints. The 8-dimensional property set are defined in **Dataset Construction** in **Section 4**, and specific values of the property sets are listed in **Table 1**.

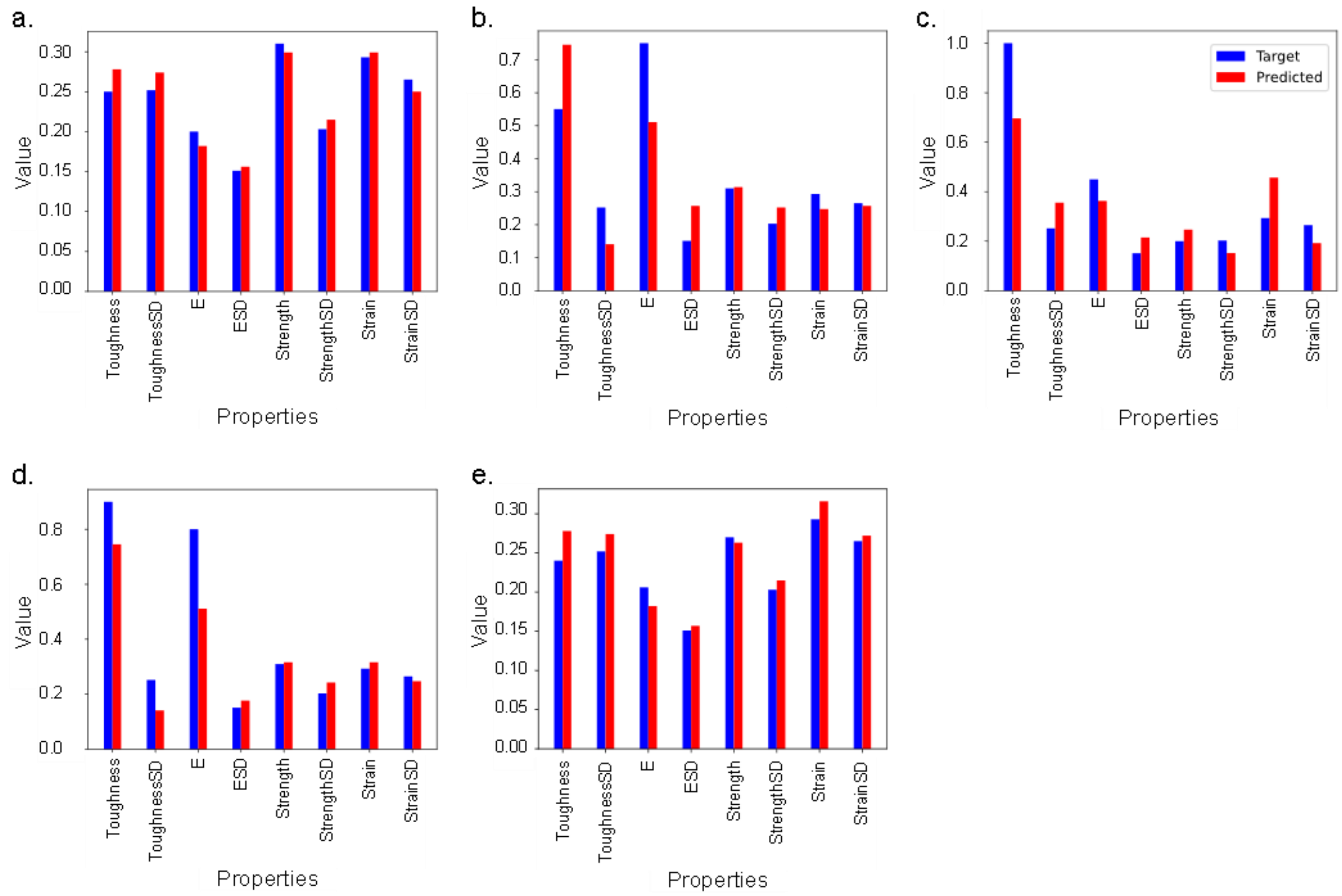

**Figure 4. Comparison of predicted values and ground truth for five selected property sets.** The bar plots present a comparison between the predicted property values (in red) of the generated novel sequences and the corresponding ground targeted properties (in blue). The generation process selects results based on the best R2 values achieved for each instance (0.8899, 0.5640, 0.7167, 0.7843, and 0.7751). Even for property sets featuring values whose combinations do not exist in nature (set 4), a comparable level of performance can be achieved (R2=0.7843). The comparison of predicted values and ground targeted properties for the five selected mechanical property sets demonstrates robust forward prediction performance of the model, even with unconventional or non-existent property values, which improves the reliability and versatility of the model for designing spider silk sequences with tailored properties. Notably, property sets that consist of common values (set 1 and set 5) tend to yield higher prediction performance, but the property sets containing rare value pairs (set 2 and set 3) get lower correlation, which suggests that the prediction capability of the model may be more adapt within the common value property combinations, and this understanding can guide the selection of property sets for specific design goals. The 8-dimensional property set are defined in **Dataset Construction in Section 4**, and specific values of the property sets are listed in **Table 1**.

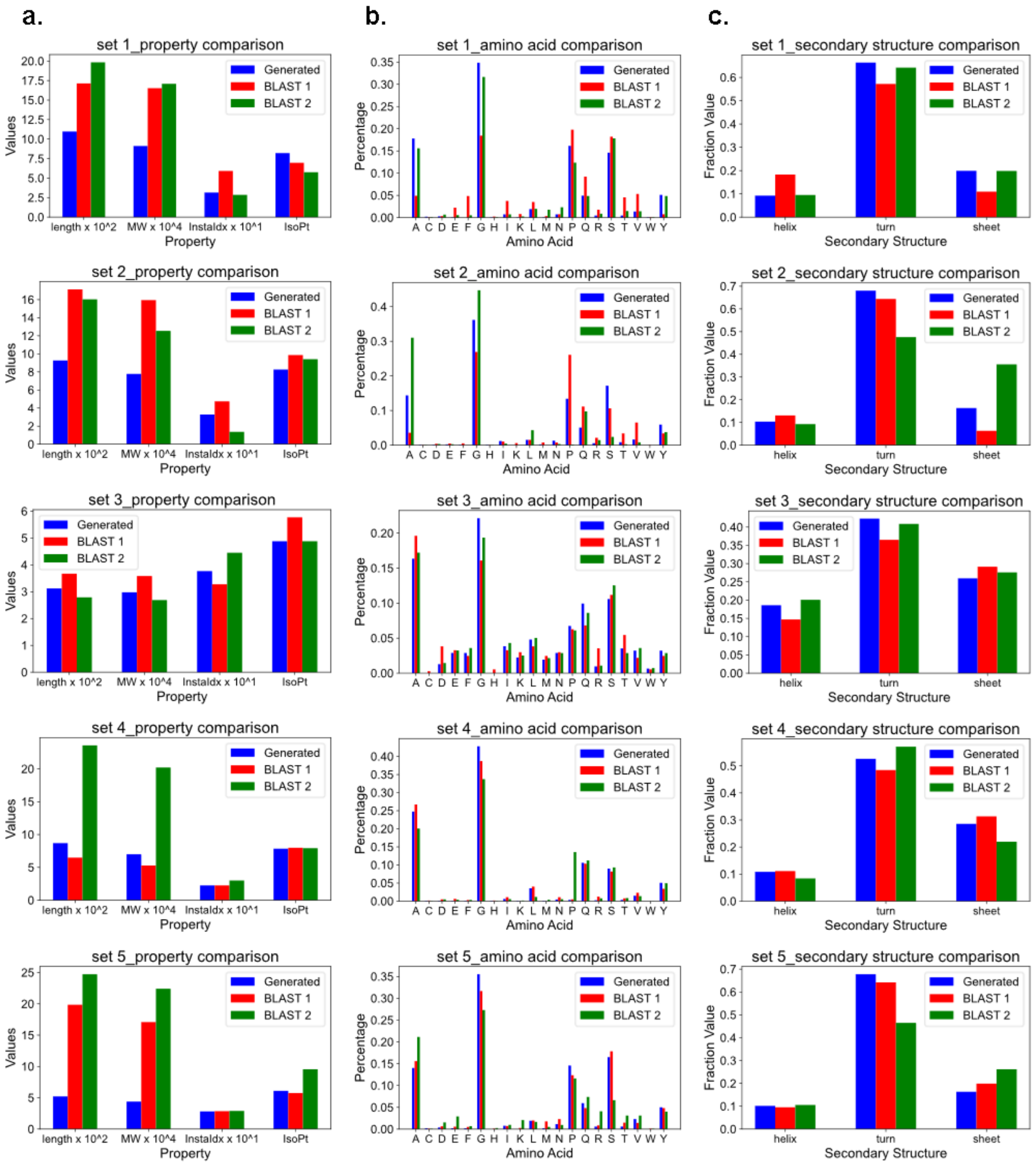

**Figure 5. Protein sequence property comparison for novel sequences (in blue) and selected sequences from BLAST search (denoted as BLAST 1 in red, and BLAST 2 in green, respectively), for five selected property sets (from top to bottom)**. The comparison include **(a)** four sequence properties: sequence length, molecular weight, instability index[89], Isoelectric point; **(b)** the amino acid percentage; **(c)** the secondary structure fraction: the ratio of amino acid tend to be in helix (V, I, Y, F, W, L), turn (N, P, G, S) and sheet (E, M, A, L)[91]. These property values are calculated using ProtParam[90] through a Python tool package[91]. Overall, a consistent trend is observed across the property values for the novel sequences and the selected BLAST searches within each property set, though value variation exists, we note that the choice of existing sequences from the BLAST searches can potentially influence the nature of the comparison. The property consistency indicates the validity of the generated sequences regarding

structural and biological features. Additionally, the consistent alignment implies that the model can effectively capture key characteristics and properties of spider silk proteins, thus reinforces the reliability of the modeling approach for novel sequence generation, and its applicability in biomaterial design.

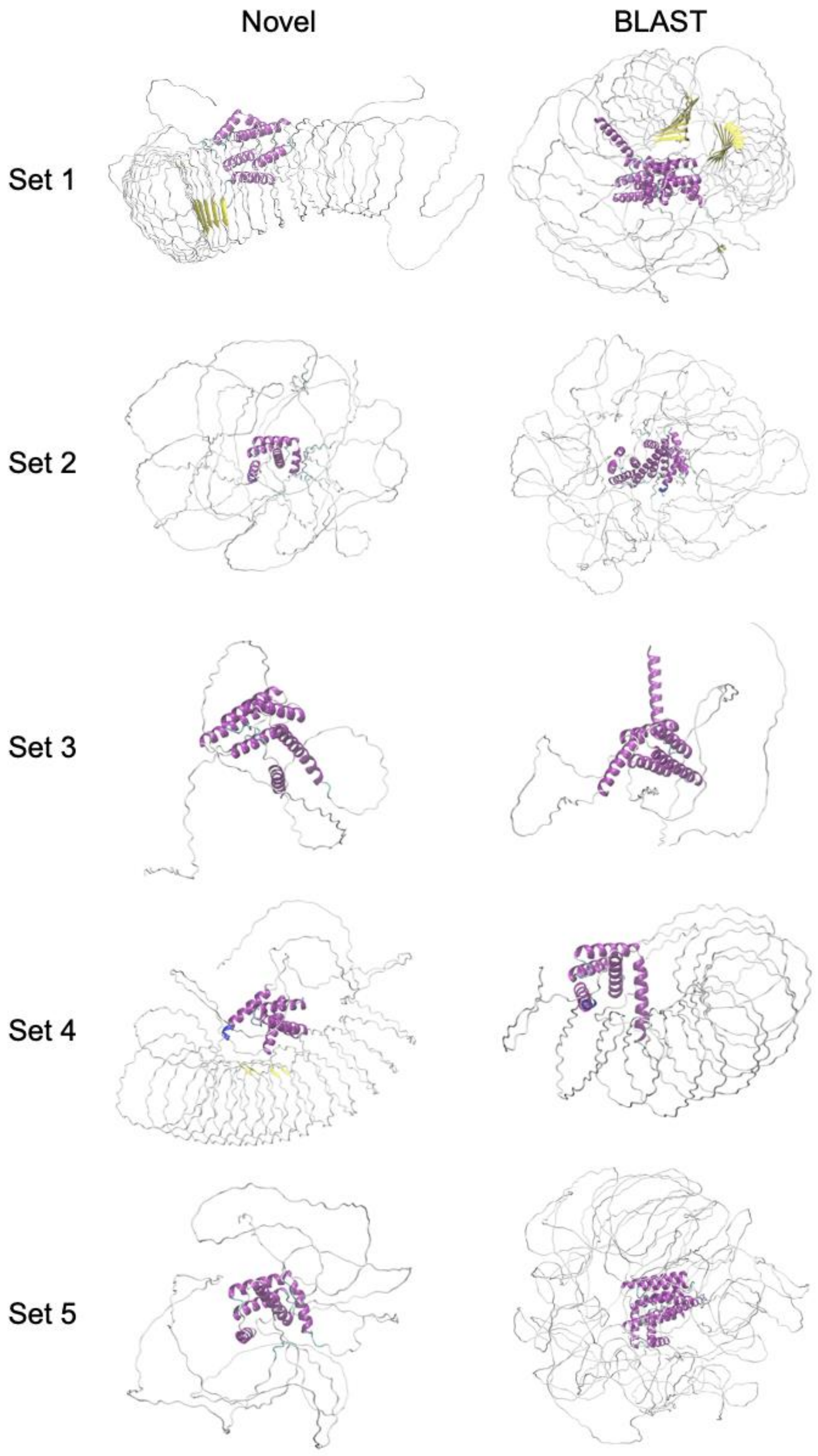

**Figure 6. Molecular configurations of spider silk sequences obtained through AlphaFold 2 (AF2)**[92], encompassing both novel sequences and 'similar' BLAST 1 sequences. Although the overall folding performance for spider silks with AF2 yields relatively low pLDDT[93,94] scores, hovering around 40 to 60 (detailed values in **Table S3**), visual similarities are observed between the configurations of novel sequences and the BLAST 1 sequences. For property sets 2 to 4, resemblances are observed regarding the alignment of alpha-helices and the overall structure and coils. While for set 1, although the overall shapes exhibit less resemblance, the presence of beta-sheets is evident in the folding of both sequences. Therefore, the accuracy of AF2 predictions for spider silk folding configurations may not be optimal, and we emphasize that the structural features should be limited to a comparative analysis between different

sequences. This may be due to the limited availability of spidroin protein sequences and the novelty of input sequences, yet the presence of visual similarities between the generated and existing sequences implies that our generation model is capable of capturing certain structural features for spider silk protein, and can generate novel and validate spidroin sequences. Furthermore, employing the generation method to compare the structural differences can potentially uncover common structural motifs or variations that may influence the properties of silks. This can enhance the understanding of the structure-property relationship of spider silks, and aid in the design of spider silks with tailored properties.

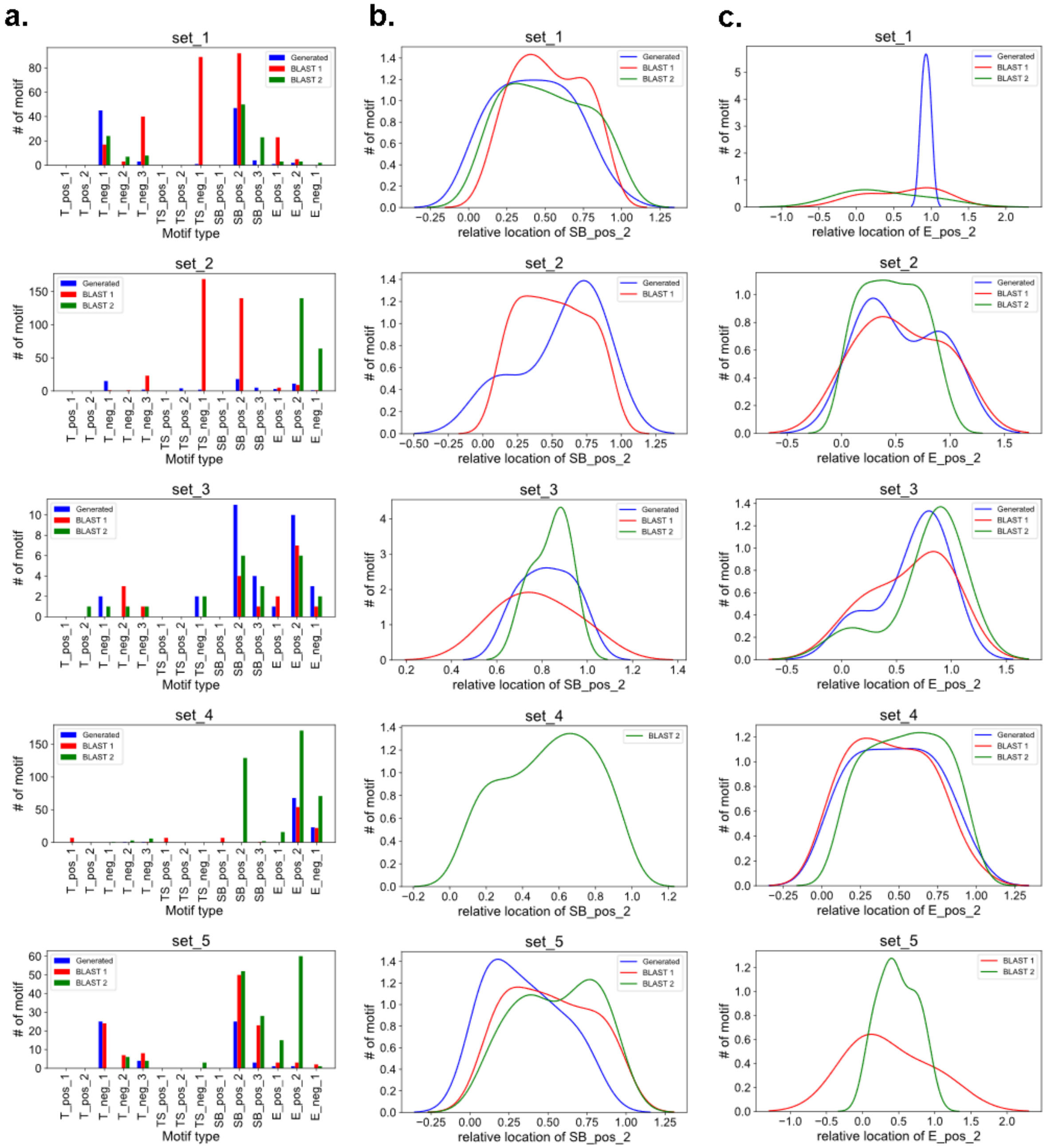

**Figure 7. Motif analysis for five mechanical property sets (from top to bottom in each series).** The bar plots in **panel a** provide comparison of motif counts for each property set. These plots depict the motif counts for both novel sequences (in blue) and two selected sequences retrieved from a BLAST search (with BLAST 1 in red and BLAST 2 in green). The motifs themselves are selected and derived from a prior study[28], with detailed information in **Table S4**. Additionally, the KDE plots in **panels b** and **c** illustrates the relative positioning pattern of specific motifs. These examples highlight the motifs "SB_pos_2" and "E_pos_2". When considering motif counts, a recurring trend of similarity is observed between novel sequences and existing ones across most property sets, particularly evident in sets 1, 3, and 4, which feature common property values and optimized yet non-existent values. The generated sequences share common structural motifs with natural spider silk sequences, which verifies their structural validity and biological relevance. The congruence is diminished in set 2 for some motifs, but note the choice of existing sequences from BLAST searches can influence this comparison. The consistency of motif impact is shown: sequences with higher toughness (sets 2, 3, and 4) display a decreased count of the "T_neg_1 & 2" motif; and sequences with elevated E values, particularly set 4, exhibit an increased count of the "E_pos_2" motif. As for the relative location of motifs, a semblance of sequence patterns is observed across most property sets. Sequences with higher E values (sets 2, 4, and also 3) demonstrate a higher level of consistency, which might be attributed to the inherent elasticity that MaSp possesses compared to other types of spider silks. The analysis validates the impact of motifs in mechanical characteristics of spider silks, as suggested in[28]; as well as demonstrates that the generative method can effectively produce sequences that align with motif impacts, encompassing both composition and positioning, thus contributing to functional properties similar to those observed in natural spider silks.

## Tables and table legends

**Table 1. Summary of selected property sets.** In this work, each protein-level sequence is paired with eight fiber-level properties for spider silk, including four mechanical properties (toughness, elastic modulus, tensile strength, strain at break), and four corresponding standard deviations derived from measurements. The eight property values are normalized and concatenated in a list for model training and prediction (details in **Section 4**). Here, three property sets are selected with random but various values, to analyze the forward prediction performance of the model (as specified in **Section 2** and **Figure 3**). Subsequently, additional five property sets, are meticulously selected (as specified **in Section 2** and **Figures 2** and **4**), to generate and analyze the novel spidroin sequences, aims to assess the self-consistency of the model, i.e. its ability to generate sequences that meet the design target.

| purpose | set # | property set values | R2 | value selection (see Figure 2) |
|---|---|---|---|---|
| 3 sets for forward prediction performance assessment | 1 | [0.600,0.204,0.279,0.043,0.329,0.127,0.502,0.139] | 0.8764 | To analysis forward prediction performance, the 3 property sets are randomly selected to encompass a range of property values. |
| | 2 | [0.300,0.20,0.2,0.1,0.3,0.1,0.20,0.1] | 0.8369 | |
| | 3 | [0.100,0.20,0.2,0.04,0.3,0.1,0.70,0.1] | 0.6889 | |
| 5 sets to generate spidroin sequences, for self-consistency analysis | 1 | [0.250, 0.252, 0.200, 0.151, 0.310, 0.203, 0.293, 0.265] | 0.8899 | Common elastic modulus and toughness value pair (0.2 & 0.25) is selected. Medians are selected for the remaining property values. |
| | 2 | [0.550, 0.252, 0.750, 0.151, 0.310, 0.203, 0.293, 0.265] | 0.5640 | Existing but rare value pair is selected for E and toughness (0.75 & 0.55), where relatively high E value is chosen. Medians are employed for other values. |
| | 3 | [1.000, 0.252, 0.450, 0.151, 0.200, 0.203, 0.293, 0.265] | 0.7167 | Existing but rare value pair is selected for E and toughness (0.45 & 1.00), where relatively high toughness value is chosen. Medians are employed for other values. |
| | 4 | [0.900, 0.252, 0.800, 0.151, 0.310, 0.203, 0.293, 0.265] | 0.7843 | Non-existent value pair for E and toughness is selected (0.8 & 0.9), and high property values are chosen. Medians are employed for for the remaining property values |
| | 5 | [0.240, 0.252, 0.206, 0.151, 0.270, 0.203, 0.293, 0.265] | 0.7751 | Common value pair of strength and toughness (0.27 & 0.24) is selected. Medians are chosen for the remaining property values. |

## References

1. Su, I. & Buehler, M. J. Nanomechanics of silk: The fundamentals of a strong, tough and versatile material. *Nanotechnology* vol. 27 Preprint at https://doi.org/10.1088/0957-4484/27/30/302001 (2016).

2. Meyers, M. A., Chen, P.-Y., Lopez, M. I., Seki, Y. & Lin, A. Y. M. Biological materials: A materials science approach. *J Mech Behav Biomed Mater* **4**, 626–657 (2011).

3. Meyers, M. A., Chen, P.-Y., Lin, A. Y.-M. & Seki, Y. Biological materials: Structure and mechanical properties. *Prog Mater Sci* **53**, 1–206 (2008).

4. Su, I. & Buehler, M. J. Mesomechanics of a three-dimensional spider web. *J Mech Phys Solids* **144**, (2020).

5. Lu, W., Lee, N. A. & Buehler, M. J. Modeling and design of heterogeneous hierarchical bioinspired spider web structures using deep learning and additive manufacturing. *Proceedings of the National Academy of Sciences* **120**, e2305273120 (2023).

6. Garrison, N. L. *et al.* Spider phylogenomics: Untangling the Spider Tree of Life. *PeerJ* **2016**, (2016).

7. Tarakanova, A. & Buehler, M. J. A materiomics approach to spider silk: Protein molecules to webs. *JOM* **64**, 214–225 (2012).

8. Su, I. & Buehler, M. J. Spider silk: Dynamic mechanics. *Nature Materials* vol. 15 1054–1055 Preprint at https://doi.org/10.1038/nmat4721 (2016).

9. Batty, L. *Molecular Dynamics Analysis of Supercontraction in Spider Dragline Silk*. (2013).

10. Liu, Y., Shao, Z. & Vollrath, F. Relationships between supercontraction and mechanical properties of spider silk. *Nat Mater* **4**, 901–905 (2005).

11. Guinea, G. V, Elices, M., Pérez-Rigueiro, J. & Plaza, G. R. Stretching of supercontracted fibers: a link between spinning and the variability of spider silk. *Journal of Experimental Biology* **208**, 25–30 (2005).

12. Cohen, N., Levin, M. & Eisenbach, C. D. On the origin of supercontraction in spider silk. *Biomacromolecules* **22**, 993–1000 (2021).

13. Boutry, C. & Blackledge, T. A. Evolution of supercontraction in spider silk: structure–function relationship from tarantulas to orb-weavers. *Journal of Experimental Biology* **213**, 3505–3514 (2010).

14. Pérez-Rigueiro, J., Elices, M. & Guinea, G. V. Controlled supercontraction tailors the tensile behaviour of spider silk. *Polymer (Guildf)* **44**, 3733–3736 (2003).

15. Buehler, E. L., Su, I. & Buehler, M. J. WebNet: A biomateriomic three-dimensional spider web neural net. *Extreme Mech Lett* **42**, (2021).

16. Yang, Q. & Li, G. Spider-silk-like shape memory polymer fiber for vibration damping. *Smart Mater Struct* **23**, 105032 (2014).

17. Kelly, S. P., Sensenig, A., Lorentz, K. A. & Blackledge, T. A. Damping capacity is evolutionarily conserved in the radial silk of orb-weaving spiders. *Zoology* **114**, 233–238 (2011).

18. Steven, E. *et al.* Physical characterization of functionalized spider silk: electronic and sensing properties. *Sci Technol Adv Mater* (2011).

19. Steven, E. *et al.* Carbon nanotubes on a spider silk scaffold. *Nat Commun* **4**, 2435 (2013).

20. Chouhan, D. *et al.* Silkworm silk matrices coated with functionalized spider silk accelerate healing of diabetic wounds. *ACS Biomater Sci Eng* **5**, 3537–3548 (2019).

21. Salehi, S., Koeck, K. & Scheibel, T. Spider silk for tissue engineering applications. *Molecules* **25**, 737 (2020).

22. Spiess, K., Lammel, A. & Scheibel, T. Recombinant spider silk proteins for applications in biomaterials. *Macromol Biosci* **10**, 998–1007 (2010).

23. Hardy, J. G., Leal-Egaña, A. & Scheibel, T. R. Engineered Spider Silk Protein-B ased Composites for Drug Delivery. *Macromol Biosci* **13**, 1431–1437 (2013).

24. Wong, J. Y. *et al.* Materials by design: Merging proteins and music. *Nano Today* **7**, 488–495 (2012).

25. Yu, C. H., Qin, Z., Martin-Martinez, F. J. & Buehler, M. J. A Self-Consistent Sonification Method to Translate Amino Acid Sequences into Musical Compositions and Application in Protein Design Using Artificial Intelligence. *ACS Nano* (2019) doi:10.1021/acsnano.9b02180.

26. Eisoldt, L., Smith, A. & Scheibel, T. Decoding the secrets of spider silk. *Materials Today* **14**, 80–86 (2011).

27. Malay, A. D., Craig, H. C., Chen, J., Oktaviani, N. A. & Numata, K. Complexity of Spider Dragline Silk. *Biomacromolecules* **23**, 1827–1840 (2022).

28. Arakawa, K. *et al.* 1000 spider silkomes: Linking sequences to silk physical properties. *Sci Adv* **8**, eabo6043 (2023).

29. Blamires, S. J., Blackledge, T. A. & Tso, I.-M. Physicochemical Property Variation in Spider Silk: Ecology, Evolution, and Synthetic Production. *Annu Rev Entomol* **62**, 443–460 (2017).

30. Babb, P. L. *et al.* The Nephila clavipes genome highlights the diversity of spider silk genes and their complex expression. *Nat Genet* **49**, 895–903 (2017).

31. Garb, J. E., Ayoub, N. A. & Hayashi, C. Y. Untangling spider silk evolution with spidroin terminal domains. *BMC Evol Biol* **10**, 1–16 (2010).

32. Arndt, T. *et al.* Spidroin N-terminal domain forms amyloid-like fibril based hydrogels and provides a protein immobilization platform. *Nat Commun* **13**, 4695 (2022).

33. Wang, S., Huang, W. & Yang, D. NMR structure note: repetitive domain of aciniform spidroin 1 from Nephila antipodiana. *J Biomol NMR* **54**, 415–420 (2012).

34. Arguelles, J. *et al.* Relating spidroin motif prevalence and periodicity to the mechanical properties of major ampullate spider silks. *Journal of Comparative Physiology B* **193**, 25–36 (2023).

35. Ayoub, N. A., Garb, J. E., Kuelbs, A. & Hayashi, C. Y. Ancient Properties of Spider Silks Revealed by the Complete Gene Sequence of the Prey-Wrapping Silk Protein (AcSp1). *Mol Biol Evol* **30**, 589–601 (2013).

36. Debabov, V. G. & Bogush, V. G. Recombinant Spidroins as the Basis for New Materials. *ACS Biomater Sci Eng* **6**, 3745–3761 (2020).

37. Savage, K. N. & Gosline, J. M. The effect of proline on the network structure of major ampullate silks as inferred from their mechanical and optical properties. *Journal of Experimental Biology* **211**, 1937–1947 (2008).

38. Huang, W. *et al.* Synergistic Integration of Experimental and Simulation Approaches for the de Novo Design of Silk-Based Materials. *Acc Chem Res* **50**, 866–876 (2017).

39. Zhang, W. *et al.* Tensan Silk-Inspired Hierarchical Fibers for Smart Textile Applications. *ACS Nano* **12**, 6968–6977 (2018).

40. Seidel, A. *et al.* Regenerated Spider Silk:  Processing, Properties, and Structure. *Macromolecules* **33**, 775–780 (2000).

41. Hinman, M. B., Jones, J. A. & Lewis, R. V. Synthetic spider silk: a modular fiber. *Trends Biotechnol* **18**, 374–379 (2000).

42. Brooks, A. E. *et al.* Properties of Synthetic Spider Silk Fibers Based on Argiope aurantia MaSp2. *Biomacromolecules* **9**, 1506–1510 (2008).

43. Petrou, G. *et al.* Genetically engineered mucoadhesive spider silk. *Biomacromolecules* **19**, 3268–3279 (2018).

44. Krishnaji, S. T. *et al.* Sequence–Structure–Property Relationships of Recombinant Spider Silk Proteins: Integration of Biopolymer Design, Processing, and Modeling. *Adv Funct Mater* **23**, 241–253 (2013).

45. Lin, S. *et al.* Predictive modelling-based design and experiments for synthesis and spinning of bioinspired silk fibres. *Nat Commun* **6**, 6892 (2015).

46. López Barreiro, D., Yeo, J., Tarakanova, A., Martin-Martinez, F. J. & Buehler, M. J. Multiscale Modeling of Silk and Silk-Based Biomaterials—A Review. *Macromolecular Bioscience* vol. 19 Preprint at https://doi.org/10.1002/mabi.201800253 (2019).

47. Gu, G. X. *et al.* Three-dimensional-printing of bio-inspired composites. *J Biomech Eng* **138**, (2016).

48. Gu, G. X., Chen, C.-T., Richmond, D. J. & Buehler, M. J. Bioinspired hierarchical composite design using machine learning: simulation, additive manufacturing, and experiment. *Mater Horiz* **5**, 939–945 (2018).

49. Milazzo, M. *et al.* 3D Printability of Silk/Hydroxyapatite Composites for Microprosthetic Applications. *ACS Biomater Sci Eng* **9**, 1285–1295 (2023).

50. Lu, W., Yang, Z. & Buehler, M. J. Rapid mechanical property prediction and de novo design of three-dimensional spider webs through graph and GraphPerceiver neural networks. *J Appl Phys* **132**, (2022).

51. Sanders, E. D., Ramos, A. S. & Paulino, G. H. Topology optimization of tension-only cable nets under finite deformations. *Structural and Multidisciplinary Optimization* **62**, 559–579 (2020).

52. Han, X. hui, Liu, H. ling, Xie, G., Sang, L. & Zhou, J. Topology optimization for spider web heat sinks for electronic cooling. *Appl Therm Eng* **195**, (2021).

53. Qin, Z., Compton, B. G., Lewis, J. A. & Buehler, M. J. Structural optimization of 3D-printed synthetic spider webs for high strength. *Nat Commun* **6**, (2015).

54. Cranford, S., Pugno, N., Vanzo, J., Cranford, S. & Buehler, M. *Simultaneous material and structural optimization in the spider web attachment disk*.

55. Chang, X. *et al.* A Comprehensive Survey of Scene Graphs: Generation and Application. (2021) doi:10.1109/TPAMI.2021.3137605.

56. Pandey, A. K. & Roy, S. S. Natural Language Generation Using Sequential Models: A Survey. *Neural Process Lett* (2023) doi:10.1007/s11063-023-11281-6.

57. Hsu, Y.-C., Yang, Z. & Buehler, M. J. Generative design, manufacturing, and molecular modeling of 3D architected materials based on natural language input. *APL Mater* **10**, 041107 (2022).

58. Buehler, M. J. Generating 3D architectured nature-inspired materials and granular media using diffusion models based on language cues. *Oxford Open Materials Science* **2**, itac010 (2022).

59. Yang, Z. & Buehler, M. J. High-Throughput Generation of 3D Graphene Metamaterials and Property Quantification Using Machine Learning. *Small Methods* **6**, 2200537 (2022).

60. Yang, Z., Hsu, Y.-C. & Buehler, M. J. Generative multiscale analysis of de novo proteome-inspired molecular structures and nanomechanical optimization using a VoxelPerceiver transformer model. *J Mech Phys Solids* **170**, 105098 (2023).

61. Ni, B., Kaplan, D. L. & Buehler, M. J. Generative design of de novo proteins based on secondary-structure constraints using an attention-based diffusion model. *Chem* **9**, 1828–1849 (2023).

62. Luu, R. K., Wysokowski, M. & Buehler, M. J. Generative discovery of de novo chemical designs using diffusion modeling and transformer deep neural networks with application to deep eutectic solvents. *Appl Phys Lett* **122**, 234103 (2023).

63. Buehler, M. J. MeLM, a generative pretrained language modeling framework that solves forward and inverse mechanics problems. *arXiv preprint arXiv:2306.17525* (2023).

64. Kipf, T. N. & Welling, M. Semi-supervised classification with graph convolutional networks. *arXiv preprint arXiv:1609.02907* (2016).

65. Jaegle, A. *et al.* Perceiver IO: A General Architecture for Structured Inputs & Outputs. in *ICLR* (2022).

66. Schleider, T. *et al.* Searching Silk Fabrics by Images Leveraging on Knowledge Graph and Domain Expert Rules. in *Proceedings of the 3rd Workshop on Structuring and Understanding of Multimedia HeritAge Contents* 41–49 (Association for Computing Machinery, 2021). doi:10.1145/3475720.3484445.

67. Dorozynski, M., Clermont, D. & Rottensteiner, F. MULTI-TASK DEEP LEARNING WITH INCOMPLETE TRAINING SAMPLES FOR THE IMAGE-BASED PREDICTION OF VARIABLES DESCRIBING SILK FABRICS. *ISPRS Annals of the Photogrammetry, Remote Sensing and Spatial Information Sciences* **IV-2/W6**, 47–54 (2019).

68. Jiang, M., Shu, T., Ye, C., Ren, J. & Ling, S. Predicting the conformations of the silk protein through deep learning. *Analyst* **146**, 2490–2498 (2021).

69. Iqbal, T. & Qureshi, S. The survey: Text generation models in deep learning. *Journal of King Saud University - Computer and Information Sciences* **34**, 2515–2528 (2022).

70. Yang, L. *et al.* Diffusion models: A comprehensive survey of methods and applications. *arXiv preprint arXiv:2209.00796* (2022).

71. Vaswani, A. *et al.* Attention Is All You Need. (2017).

72. Devlin, J., Chang, M.-W., Lee, K. & Toutanova, K. Bert: Pre-training of deep bidirectional transformers for language understanding. *arXiv preprint arXiv:1810.04805* (2018).

73. Brown, T. *et al.* Language models are few-shot learners. *Adv Neural Inf Process Syst* **33**, 1877–1901 (2020).

74. Black, S. *et al.* Gpt-neox-20b: An open-source autoregressive language model. *arXiv preprint arXiv:2204.06745* (2022).

75. Biderman, S. *et al.* Pythia: A suite for analyzing large language models across training and scaling. in *International Conference on Machine Learning* 2397–2430 (PMLR, 2023).

76. Lan, Z. *et al.* Albert: A lite bert for self-supervised learning of language representations. *arXiv preprint arXiv:1909.11942* (2019).

77. Wang, B. and K. A. GPT-J-6B: A 6 Billion Parameter Autoregressive Language Model. https://github.com/kingoflolz/mesh-transformer-jax (2021).

78. Kluge, J. A., Rabotyagova, O., Leisk, G. G. & Kaplan, D. L. Spider silks and their applications. *Trends Biotechnol* **26**, 244–251 (2008).

79. Liu, Y. *et al.* Highly flexible and lightweight organic solar cells on biocompatible silk fibroin. *ACS Appl Mater Interfaces* **6**, 20670–20675 (2014).

80. Yodmuang, S. *et al.* Silk microfiber-reinforced silk hydrogel composites for functional cartilage tissue repair. *Acta Biomater* **11**, 27–36 (2015).

81. Sun, W., Gregory, D. A., Tomeh, M. A. & Zhao, X. Silk fibroin as a functional biomaterial for tissue engineering. *Int J Mol Sci* **22**, 1499 (2021).

82. Kundu, B. & Kundu, S. C. Silk sericin/polyacrylamide in situ forming hydrogels for dermal reconstruction. *Biomaterials* **33**, 7456–7467 (2012).

83. Farokhi, M., Mottaghitalab, F., Fatahi, Y., Khademhosseini, A. & Kaplan, D. L. Overview of silk fibroin use in wound dressings. *Trends Biotechnol* **36**, 907–922 (2018).

84. Yucel, T., Lovett, M. L. & Kaplan, D. L. Silk-based biomaterials for sustained drug delivery. *Journal of Controlled Release* **190**, 381–397 (2014).

85. Lammel, A. S., Hu, X., Park, S.-H., Kaplan, D. L. & Scheibel, T. R. Controlling silk fibroin particle features for drug delivery. *Biomaterials* **31**, 4583–4591 (2010).

86. An, B. *et al.* Reproducing natural spider silks' copolymer behavior in synthetic silk mimics. *Biomacromolecules* **13**, 3938–3948 (2012).

87. Kono, N. & Arakawa, K. Nanopore sequencing: Review of potential applications in functional genomics. *Dev Growth Differ* **61**, 316–326 (2019).

88. Quan, L., Wu, T. & Lyu, Q. Computational de novo protein design: From secondary to primary, then toward tertiary structures. *Chem* **9**, 1625–1627 (2023).

89. Guruprasad, K., Reddy, B. V. B. & Pandit, M. W. Correlation between stability of a protein and its dipeptide composition: a novel approach for predicting in vivo stability of a protein from its primary sequence. *Protein Engineering, Design and Selection* **4**, 155–161 (1990).

90. Gasteiger, E. *et al.* Protein Identification and Analysis Tools on the Expasy Server; in *The Proteomics Protocols Handbook, Humana Press* 571–607 (2005).

91. Official git repository for Biopython: ProtParam.py. *Github*.

92. Jumper, J. *et al.* Highly accurate protein structure prediction with AlphaFold. *Nature* **596**, 583–589 (2021).

93. Guo, H.-B. *et al.* AlphaFold2 models indicate that protein sequence determines both structure and dynamics. *Sci Rep* **12**, 10696 (2022).

94. Varadi, M. *et al.* AlphaFold Protein Structure Database: massively expanding the structural coverage of protein-sequence space with high-accuracy models. *Nucleic Acids Res* **50**, D439–D444 (2022).

95. Paszke, A. *et al.* PyTorch: An Imperative Style, High-Performance Deep Learning Library. (2019).

96. Kingma, D. P. & Ba, J. Adam: A Method for Stochastic Optimization. (2014).

# Generative modeling, design and analysis of spider silk protein sequences for enhanced mechanical properties

Wei Lu[1,2], David L. Kaplan[3], Markus J. Buehler[1,2,4*]

[1] Laboratory for Atomistic and Molecular Mechanics (LAMM), Massachusetts Institute of Technology, 77 Massachusetts Ave., Cambridge, MA 02139, USA

[2] Department of Civil and Environmental Engineering, Massachusetts Institute of Technology, 77 Massachusetts Ave., Cambridge, MA 02139, USA

[3] Tufts University, Medford, MA

[4] Center for Computational Science and Engineering, Schwarzman College of Computing, Massachusetts Institute of Technology, 77 Massachusetts Ave., Cambridge, MA 02139, USA

* email: mbuehler@mit.edu

# SUPPLEMENTARY MATERIALS

**Supplementary figures**

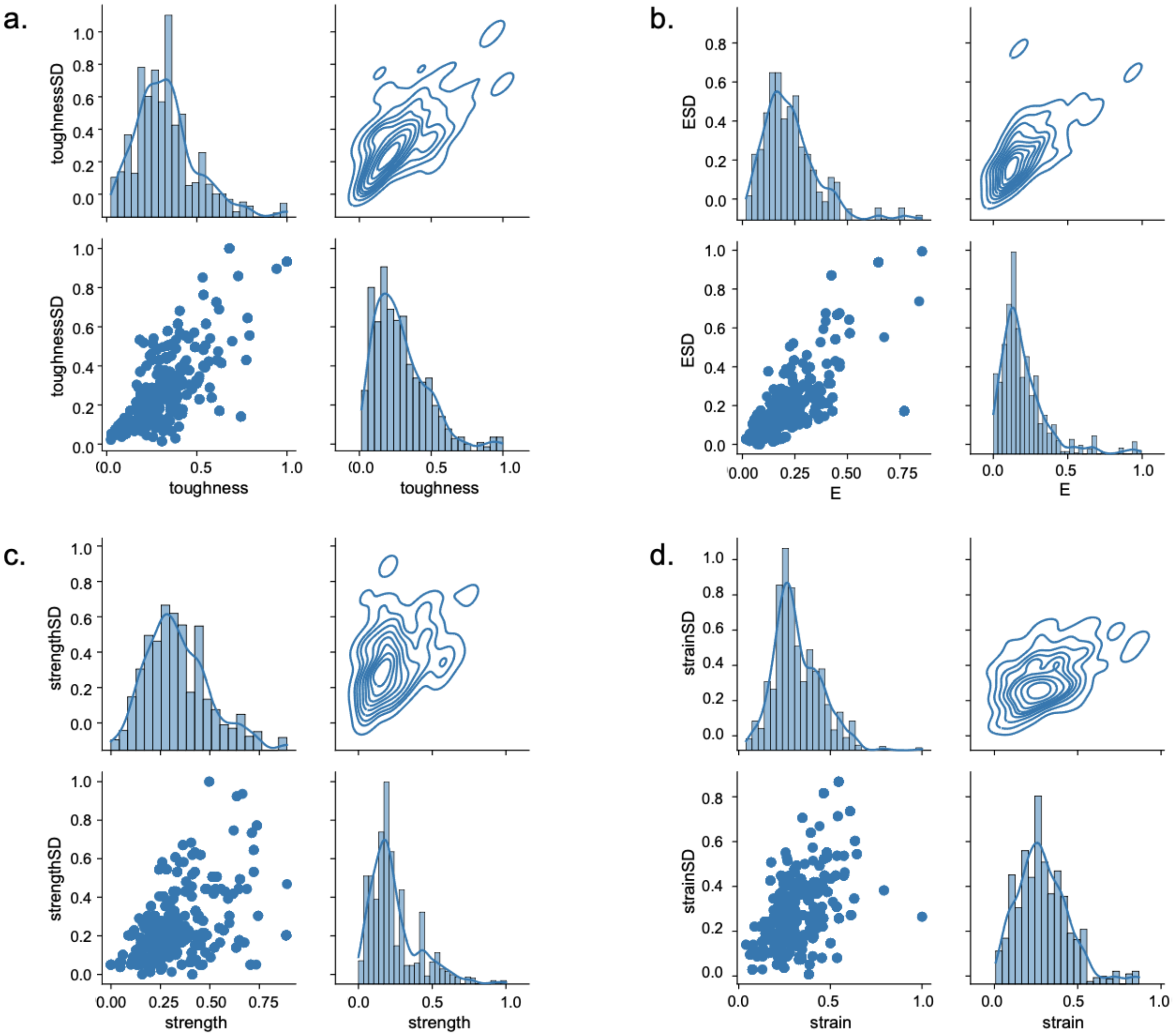

**Figure S1. Analysis of the normalized mechanical properties and their corresponding standard deviations** within the MaSp silk dataset from Silkome[1], for **(a)** toughness, **(b)** elastic modulus, **(c)** tensile strength, and **(d)** strain at break. The distribution of data is depicted through histograms with kernel density estimation (kde) plots on the diagonal within each panel. Additionally, the correlation between properties and their standard deviations (SDs) is visualized via scatter plots in the lower region and kde plots in the upper region. Uniform distributions are observed for each property and their respective standard deviation values. The toughness and elastic modulus are more positively correlated with their standard deviation values, than strength and strain. The variability or uncertainty in measurements tends to increase when the properties increase, especially toughness and elastic modulus, which may be due to certain factors affecting the spider silk property or the measurement process as the values increase.

## Supplementary tables

**Table S1. Summary of generated spider silk sequences.** Five generated novel spider silk sequences, and ten similar existing spider silk sequences from BLAST search, with two for each generated sequences (designated as BLAST 1 and BLAST 2), are summarized here. The sources of ten existing sequences are indicated in each row. The selection criteria for the existing sequences involve relatively high query cover and percent identity, as well as comparable sequence length when contrasted with the generated sequences.

| Type | ID | Sequence | NCBI |
|---|---|---|---|
| Generated | 1.1 | GPGSQGPSGPGGYGPGSQGPSGPGGYGPGSQGPSGPGGAAAAAAAAASGPGGYGPGSQGPSGPG<br>GYGPGSQGLSGLGGAAAAAAAAASGPGGYGPGSQGPSGPGGYGPRNLQQGPSGPGGAGAAAAAA<br>AASGPGGYGPGSQGPSGPGGYGPGSQGPSGPGGAAAAAAAAASGPGGYGPGSQGPSGPGGYGPG<br>SQGPSGPGGAAAAAAAAASGPGGYGPGSQGPSGPGGYGPGSQGPSGPGGAAAAAAAAASGPGGYG<br>PGSQGPSGPGGYGPGSQGPSGPGGAAAAAAAAASGPGGYGPGSQGPSGPGGYGPGSQGPSGPGG<br>AAAAAAAASGPGGYGPGSQGPSGPGGYGPGSQGPSGPGGAAAAAAAAASGPGGYGPGSQGPSGP<br>GGYGPGSQGPSGPGGAAAAAAAAASGPGGYGPGSQGPSGPGGYGPGSQGPSGPGGAAAAAAAAS<br>GPGGYGPGSQGPSGPGGYGPGSQGPSGPGGAAAAAAAAASGPGGYGPGSQGPSGPGGYGPGSQG<br>PSGPGGAAAAAAAAASGPGGYGPGSQGPSGPGGYGPGSQGPSGPGGAAAAAAAAASGPGGYGPGS<br>QGPSGPGGYGPGSQGPSGPGGAAAAAAAAASGPGGYGPGSQGPSGPGGYGPGSQGPSGPGGAAA<br>AAAAASGPGGYGPGSQGPSGPGGYGPGSQGPSGPGGAAAAAAAAASGPGGYGPGSQGPSGPGGY<br>GPGSQGPSGPGGAAAAAAAAASGPGGYGPGSQGPSGPGGYGPGSQGPSGPGGAAAAAAAAASGPG<br>GYGPGSQGPSGPGGYGPGSQGPSGPGGAAAAAAAAASGPGGYGPGSQGPSGPGGYGPGSQGPSG<br>PGGYGPGASGPGGYLGPGGYGPGSQGPGGPGASAAAAAASGPGGYGPGSQGPSGPGYQGPSGP<br>GAYGPSPAASASVAASRLSSPQASSRVSSAVSSLVGPSGPGASGSSVSSALNSLTSQISSRNP<br>GPSGPGGYGPCDGDGSSVVSSLVTILSSASIGQVNYGSGGYGPGSQGPGGASAAISSALNSLT<br>MGWLTNIGPSGPGGYYKTAAAASAAASRLSSPGICDVLVQALLEVVSALVHILSSASIGQVNY<br>GSGGYGPGSQGPSGPGGYGPNLAGL | N/A |
| | 2.1 | GPGSQGPSGPGGAGGYGPGSQGPGGASAAAAAAAAGGPGGYGPGSQSGPGGYGPGSSGPGGAGG<br>YGPGSSGPGGAGGYGPGSQGQGGSGAAAAAAAAASGPGGYGPGSQGPSGPGSSGPGGAGGYGPG<br>SSSSVISTALAGGYGPGSSGPGGAGGYGPGSSGPGGAGGYGPGSSGPGGAGGYGPGSQGQGGS<br>GAAAAAAAASGPGGYGPGSQGPSGPGSSGPGGAGGYGPGSSGPGGAGGYGPGSSGPGGAGGYG<br>PGSSGPGGAGGYGPGSQGQGGSGAAAAAAAAASGPGGYGPGSQGQSGPGSAGQIPGPGSGQQGP<br>AGYGPSGPSGPGGYGPGSSASAAAAAAASGPGGYGPNNGYGPGGAGGYGPGSSAAAAVAAAGGY<br>GPGAIGYREALSNAMSNPYSQSSRMSSAAAAAAAAASGPGGYGPGSSGPGGAGGYGPGSSAIYG<br>PGYQIGLYGPGSAPQQTISQTGNTSALPTISQSSGPSAATGGYGPGSSGPGGYGPGSQGPSGP<br>GGAPGTRPGAYGPSSQGPSGPGGYGPGSSAAAAAAAASGPGGYGPGSQGPSGPGSQGPSGPGGY<br>GPGSSGPGGPAGGYGPGSSGPGGAGGYGPGSSGPGGAGGYGPGSQGQGGSGASAAAAAAASGPG<br>GYGPGSQGPSGPGSQGPSGPGGYGPGASGPGGAGGYGPGSSGPGSGAGGYGPGSSASAAAAAA<br>SASGPGGYGPGSQGPSGPGSQGPSGPGGYGPGSQGPSGPGGYGPGASGPGGAGGYGPGSQGAG<br>GGGPGGAGGYGPGNQGSGGQQGPGAYGPGSQGPSGSSAAAAASSGGYGPGSQGPSGYYGPVQQ<br>GPSSSNPASAAAARLSSPQASSRVSSAVSSLVSSGPSNGAALSNTISNVVSQISSSNPGLSGC<br>DVFVQALLELVSALVHILSSSSIGQVNYGASGQYASLVGQSVAY | |
| | 3.1 | MNWSIRLALLGFVVLSTQTVFAVGQAATPWENSQLAEDFINSFLRFIGQSGAFSANQLDDMSS<br>IGDTLKTAIEKMAQSRKSSKSKLQALNMAFASSMAEIAVAEQGGLSLEAKTNAIANALTSAFL<br>ETTGVINQQFVSEIKGLIYMIAQASSNEISGSSAAAGGGSGGGYGGGGSGQGGYGQGAYASA<br>SAAVAYGSAPQGAGGPAPQGPSQQGPVSQGPYGPSAAAAAAATGGYGPGAGQQGPGAGQQGPG<br>SGGQQGPGGQGPYGPGAAAAAAAVGGYGPGAGQQGPGSQGPSGPGGTQQGPGGQGPYGPG | |
| | 4.1 | AAAAGAGGQGGYGGLGSQGAGQGGYGAGQGGAASAAAGGAGGQGGYGAGQGGAFSSSISAAAG<br>GAAQYKQTSGAAASAAGGAGGQGGYGGLGSQGAGQGGYGAGQGGAASAAAGGAGGQGGYGGLG<br>SQGAGQGGYGAGQGGAGAAAAAAAAAAGGAGGAGQGGLGAGQGYGSGLGGQGGAGQGGYGAGQG<br>GAASAAAAAGGSGGQGGYGGLGSQGAGQGGYGAGQGGAGSAAAAGGAGGQGGYGGLGSQGAGQ<br>GGYGAGQGGAAASSAAASAAGGAGGQGGYGGLGSQGAGQGGYGAGQGGAASAAAGGAGGQGGY<br>GGLGSQGAGQGGYGAGQGGAGAAAAAGGAGGQGGYGGLGSQGAGQGGYGAGQGGAASAAAGGA<br>GGQGGYGGLGSQGAGQGGYGAGQGGASAAAAAGGAGGQGGYGGLGSQGAGQGGYGAGQGGAAA<br>SAAAGGAGGQGGYGGLGSQGAGQGGYGAGQGGASAAAGGAGGQGGYGGLGSQGAGQGGYGAGQ<br>GGAGSSAAAAAAAGGAGGQGGYGGLGSQGAGQGGYGAGQGGASAAAAGGAGGQGGYGGLGSQG<br>AGQGGYGGQYGAGQGGAAAAAAAAAAAGGAGGQGGYGGLGSQGAGQGGYGAGQGGAGAAAAAGG<br>AGGQGGYGGLGSQGAGQGGYGAGQGGASAAAGGAGGQGGYGGLGSQGAGQGGYGAGQGGAASA<br>AAAAGGAGGQGGYGGLGSQGAGQGGYGAGQGGASAAAAGGAGGQGGYGGLGSQGAGQGGYGGQ<br>GAAASSAAASAAASRLSSPSAASRVSSAVSSLVSSGGPTSSSALSSTISNVVSQVSASNPGLS<br>GCDVLVQALLEIVSALVHILGSANIGQVNSSAAGQSASMVGQSVSQALSY | |
| | 5.1 | GGPGGYGPGSQGPSGPGGAAAAAAAAAGGPGGYGPGSQGPGSQGPGSQGPGSQGPGGYGPGSQG<br>PGGPGGYGPGSQGPGSQGPSGPGGYGPGSQGPSGPGGAAAAAAAAAAGGPGGYGPGSQGPSGSR<br>GPSGPGGYGPGSQGPSGPGGYGPGSQGPSGPGGAGGYGPGSQGPGGASAAAAAASGPGGYGPG<br>SQGPSGPGSQGPSGPGGYGPGSQGPSGPGGYGPGASGPGGAGGYGPGSQGPGGASAAAAAAAA<br>ASGPGGYGPGSQGPSGPGSQGPSGPGGYGPGSQGPSGPGGYGPGASGPGGAGGYGPGNQGSGG<br>ASAAAAAAAASGPGGYGPGSQGPSGPGSQGPSGPGGYGPGASGPGGAGGYGPGSQGPGGASAA<br>AAAAASGPGGYGPGSQGPSGSGAYGPGSQGPSGYGGYGPSASASVSSAASRLSSPAASSRVSS<br>AVSSLVSNGASNGASVSGALNGLVSQISSSNPGLSGCDVLVQALMELVSALVAILGSASIGSV<br>DYNSVGQTTQTISQYFS | |
| BLAST1 | 1.2 | MNWSKTFSLLCLLVLSTQALLLVEAARSPWESTQLAESFLKSFLRAIARSGAFSSNQLDDMST<br>IGETLTTSIEKLASSSKTSKAKLQALDMAFASSMAEIAVAEEGGLSINEKTEAIGNALKTAFL<br>ETTGRINAQFVSEIKSLIFLIAQATSNEITSASPTGATGYGTPGQGGTYASVSIAGSYSQTPQ<br>RPQTTGQEPSSQGPLSPEPQGTAFSSVSSYGPGPQGPSGPGPQTPLPQGPSGPGLQISGPSVG<br>ILSSGPGPQGSSGPIPQESFPEGSSSPATQGPSVSSISFFEPGPKGPGGPGPQAPLPQGPSGP<br>GPQRPTFSSVSAYGPGSLGPSGPGPQAPSPQGPSGPGPQGPTVSSISFFGPGPQAPSLQGPSG<br>LGTQGPEQSASVLTSYGPGPQGSLKVPSAQGPYIPKSQGPAVSSVSFFGPGRQGSSRPSPQGP<br>IPQGPSGPGPQGPVARSVSFFGPGTQGPTGPSPEGPITQGPSVPGPQGSAVSSVSFFGPGSQG<br>PSGPSTQGPIPQGPSGPGPQGPVARSVSFFGPGTQGPTGPSPEGPITQGPSVPGPQGSVVSSV<br>SFFGPGSQGPSGPSTQGPIPQGTSGTGPQGPAVSSVSFFGPGTRGPSGPSPQGTIPQGPSGPE<br>PQGPAASSVSFFRPGTQGPSGPSPQGPIPQGPSVPGPQGSAVSSVSFFIPGSQGPSGPSTQGP<br>IPQGPSSPGTQGPAVRSVSFESGSQGPYGPSPQGPIPQRPSGPGSQVPTESSVSFEPGSQGPY<br>ESSPQGPIPQHPSGPGPQGPPASSVSFFGPGTQGPSGPSPQGPIPQGTSGTGPQGPAVSSVSF<br>FGPGPQGPSGAIPQGPSGPGTQVPTESSVSFEPGSQGPYESSPQGPIPQRPSGPGPQGPPASS<br>ISFFGPGTQGPSGPSPQGPIPQGPSGTGPQGPAVSTVSFFGPGPQEPSGSIPLGPSGPGPQGP<br>PASSVSFFGPGTQGPSGPSPQGPIPQGSSVPGSQGSAVSSVSFFGPGSEGPSGPSTQGPIPQG | UZZ64713.1 |
| | | TSLTGPQGPAVSSVSFFGPGPLGPSGPITQRPTSQGPSGPVPQGPAVSFVSFFAPGPQGPSGP<br>SLQGPIPQGTSGPGPQGSSLSSVSFFGPGSQGPSGPSSHGPIPQGPFGPGSQGPTVSTVSFFG<br>LGPQGPFGPSPQGSIPQGPSIPGPQVPAVSSVSFFEPGPQGLSGPSPQGPIPQGPSIPGPQVP<br>AVSSVSFFGPGSQGPSGPSTQGSIPQGTSGTGPQGPAVTSVSFEPGSQRPYGPSPQGPIPQGP<br>SGPGPHGPAASSVSFFGPGTQGPSGPSPQGPIPQGPFGPGSQGPTVSSVSFFGPGSQGPSGPS<br>TQGPIPQGTSGTGPQGPAVSSVSFFGPGTRGPSGPSPQDPIPQGPSIAGPQVPAVSSVSFFGP<br>GSQGPSEPSPQGPISRGPSGPGPQGPAVSSVSFFGPRPRGPSGPSAQGPIPQGPSFPGPQRPA<br>VSSVSFEPGSQGPYGPSPPGSILQGPSGAGPQGPAASSVSFFGPGTQGPSGPSPQGPIPQGPF<br>GPGSQGPTVSSVSFFGSRPQGPSGPSPVGPIPQGLSGPLPQAPAVSSLSMYDAVDQGLSRFAS<br>QVPLPLGQSARTPQRSSASFGLGLSGVSVLHQSPVVSSAASRLSSPAAISRISSAMSSLAASG<br>PRNPIGLSKVLGSISSQIKASNPGLSECETFAQTLLEIVSALIQILNSSNIGQVNLRPTGQSN<br>AVVNQAVLQTLG | |
| | 2.2 | MSYLTRLALALLAVLSTQAIFANGHITPWSNTRLAEAFINSFMSKVGYSGAFTAEQMDDMSTV<br>SDTIMTAMDKMASSNKSSKSKLQALNMAFASTMAEIAATEEGGQSMSVKTNAITDALSAAFLE<br>TTGQVNYQFINEIKSLVYMLAQQSMNDVYASAGTASGGGPGPQGPAPQGPSGPGPQRPQGPGP<br>YGPSGASVFSATVIGPGPQGPSGPGPQGPYGSGPSGPGPQGLAPQGPSGPSPQRPQGPGPQGP<br>YGPGGVSVVSATVSGPGPQGPSGPGPQGPYGPRPQGPGPQGPGPQGPSGPGPQGPSGPRPQGP<br>YGPGGVSVVSTTVSSPGPQGPSGPGPQGPYGPGPQGPGPQGPGPQGPSGPGPQGPSGPGPQGP<br>YGPGGVSVVYTTVSGPGPQGPSGPGPQGPYGPRPQGPGPQGPGPQGPSGPGPQGPSGPRPQGP<br>YGPGGVSVVSTTVSSPGPQGPSGPGPQGPYGPGPQGPGPQGPGPQGPSGPGPQGPSGPGPQGP<br>YGPGGVSVVYTTVSGPGPQGPSGPGPQGPYGPGPQGPGPQGPAPQGPSGPGPQRPQVPGPQRP<br>YGSGGVSVVSGTVSGPGPQGPSGPGPQGPYGPRPQGPGPQGQGPQGPGPQRPQGPGPQGPYGP<br>GGVSVVSTTVSSAGPQRPSGPGPQGPYGPGQQGPGPQGPGPQGPGPQGPSGPGPQGPYGPGGV<br>SVVSTTVSGSGPQGPSGPGPQGPYGPRPQGPGPQGPSPQGPSGPGPQRPQGPGPQSPYSPGGV<br>SVVYTKVSGPGPQGPSGPGPQGPYGPRPQGPGPQGQGPQGPRPQGPSGPGPQGPYGPGGVSVV<br>STTVSSPGPQGPSGPGPQGPYGPGPQGPGPQGRGPQGPGPQGPSAPGPQGPYGPGGVSVVSTT<br>VSGSGPQVPSGPGPQGPYGPGPQGPSPQGPSGPGPQRPQGPGPQGPYGPGGVSVVYTTVSGPG<br>PQGPSGPGPQVPYGPGPQGPGPQGPGPQRPVPQRPQGPGPQGPYGAGGVSVVSATVSKPGPQG<br>PSGPGPQGPYGPRPQGPGPQGQGPQGPGPQGPSGPGPQGPYGPGGVSVVSTTVSSPGPQGPSG<br>PGPQGPYGPGPQGPGPQRPGSQGPGPQGPSGPGPQGPYGPGGVSVVSTTVSGSGPQGPSGPGP<br>QGPYGPGPQGPGPQRPSPQGPSGPGPQRPQGPGPQGPYGPGVISVSAVSGPGPQGPSGPGPQV<br>PYGPRPQGPGPQGPGPQGPVPQRPQAPGPQVPYGAGGVSVVSATVSKPGPQGPSGPGPQSPYG<br>PGPQGPGPQGPTPQGPSGPGPQRPQGPGPQGPYGPGGVSVVSATISGPGPQGTSGPGPQGPYG<br>PGPQGPGPQGPGPQGPTGPGPQVPSGPGPQGPYGPSGVSVVSTTVSGPGPLGPSGPAPQGPYG<br>PGPQGPGPRGPTPQGPSGPGPQRPQGPGPQGPYGPGGVSVVSATISGPGPQGPSGPGPQGPYG<br>PGPQGPGPQGPGPQGPTGPGPQGPSGPGPQGPYGPGGVSVVSTTVSGPGPQGPSGPGPQGPYG<br>PGPQGPGPQGPAPQGPQGPGPQRPQGPYGPGGVSLVSATVSGPGPQGPSGPGPQGPYGPRPQG<br>PGPQGAGPGPQRPSGPGPQGPYGPGPRSPSSTPGSAAINAASRLSSPAASSRVSSTVSQLVSS<br>GTPNSAAVSGAISSLVSQVSASNPGLSGCDILVQALMELLSALVSIVGSSSIGQVNYGASGQY<br>AQLVSQAIGQAF | GIX92750.1<br>from[2] |
| | 3.2 | MSWSTLALAILVVFSTQCIFIAGQANTPWSDTATADAFIQNFLGAVSGSEAFTPDQLDDMATV<br>GDTIMSAIDKMARNNKSSKSKLQSLKMAFASSIAEIAAVEQGGQSMDIKTNAIANALDSAFYM<br>TTGSTNQQFVNEMRRLINMLSAASVNELSYGGGTSAAAATAGGYGQGASGYDPGSSPASAAAL<br>KGPSKREPSGLGAAAAASSEYGSSQQGPSGTKAATIAAAKRGPTSYGPRQQRPGGSGAPAATA<br>GRGPGSYGPEQQGPRGSGAAADEAEPGQQEPGADAAAALGSGSGDQGPGRFDAAAATAKPRGN<br>GPGQQGSGVASAAAAGSGPRGYGPGQQAHRGHGAAAAATGSGGYEPGQQGPG | AWK58654.1<br>from[3] |
| | 4.2 | AAAAAAAAAGGQGGQGGYDGLGSQGAGQGGYGQGGAAAAAAAAASGAGSAQRGGLGAGGAGQGY<br>GAGSGGQGGAGQGGAAAATAAAAGGQGGQGGYGGLGSQGSGQGGYGQGGAAAAAAAAASGDGGA<br>GQEGLGAGGAGQGYGAGLGGQGGAGQGGAAAAAAAAAAGGQGGQGGYGGLGSQGAGQGGYGQGG<br>AAAAAAAASGAGGAGQGGLGAAGAGQGYGAGSGGQGGAGQGGAAAAAAAAAAGGQGGQGGYGGL<br>GSQGAGQGGYGQGGVAAAAAAAASGAGGAGRGGLGAGGAGQEYGAVSGGQGGAGQGGEAAAAAA<br>AAGGQGGQGGYGGLGSQGAGQGGYGQGGAAAAAAAAASGAGGARRGGLGAGGAGQGYGAGLGGQ<br>GGAGQGSASAAAAAAAAGGQGGQGGYGGLGSQGSGQGGYGQGGAAAAAAAAASGAGGAGRGSLGA<br>GGAGQGYGAGLGGQGGAGQGGAAAAASAAAGGQGGQGGYGGLGSQGAGQGGYGQGGAAAAAAS<br>AGGQGGQGGYGGLGSQGAGQGGYGGGAFSGQQGGAASVATASAAASRLSSPGAASRVSSAVTS<br>LVSSGGPTNSAALSNTISNVVSQISSSNPGLSGCDVLVQALLEIVSALVHILGSANIGQVNSS<br>GVGRSASIVGQSINQAFS | AAK30595.1<br>from[4] |
| | 5.2 | MSYPRLVLAFLALLSTHALFAAAGGQTPWDTPTLADNFMKCFMNEIGNSGAFTSNQVDDMSTI<br>GDTMMDSVNRLASSGRISKSKLQALNMAFASSMAEIAATEEGGLSIGSKTNAIASALRGAFLQ<br>TTGYSNEQFINEITSLVSMIAEANVNTVSASASAYAGGGYGGSSYGSSSVNSASAAATGPAQQ<br>GPGSYGPSSVPGGYGPSGSSAAAAASGGQGLGNYGPSGSGGAGPSGTGGYGPGSQGPSRPSGP<br>GAAAAAAAASGPGGYGPSAPSGPGSQGPSGPSGSGASAAAAAASGSGGYGSGSQGPSGPGSQG<br>PSGPGTSAAAAAAANGPGGYGPGSQGPSGPGGYRPGSQGPSGPGSSGPGMSGGYGPGNQGPGG<br>ASVAAAAAGSGPGGYGPGSQGPSGPGSSGPGMSGGYGPGNQGPGGASAAAAAASGPGGYGPGS<br>QGSSGPGAYGPGSQGSSGPGSSGPGMSGGYGPGNQGPGGASAAAAAAAASGPGGYGPXSKGPGR<br>HXGPGSSGPGMSGGYGPGNQGPGGASAAAAAAAASGPGGYGPGSQGPSGPGAYGPGSQGSSGPG<br>SSGPGMSGGYGPGNQGPGGASAAAAAAAASGPGGYGPGSQGPSGPGAYGPGSQGSSGPLSSGPD<br>MSGGYGPGNQGPGGASAAAAAAAASGPGGYGPGSQGPSGPGAYGPGSQGSSGPGSSGPGMSGGY<br>GPGNQGPGGASAAAAAVASGPGGYGPGSQGSSGTGAYGTGSQGSTGPGSSGPGMSGGYGPGNQ<br>GPGGASAAAAAAAASGPGGYGPGSQGSSGPGAYGPGSQGSSRPLSSGPGMSGGYGPGNQGPGGA<br>SAAAAAAASGPGGYGPGSQGPSGPGAYGSGSQGSSGPGSSGPGMSGGYGPGNQGPGRASAAAA<br>AAASGPGGYGPGSQGPSGPGAYGPGSQGSSGPLSYGTDMSGGYGPGNQGPGGASAAAAAAASG<br>PGGYGPGSQGPSGPGAYGPGSQGSSGPGSSGPGMSGGYGPGNQGPGGASAAAAAEASGPGAYG<br>PGSQGSSRLLSSGPGMSGGYGPGNQGPGGASAAAYAAASGLGGYGPGSQGPSGPGAYGSGSQG<br>SSGPGSSGPGMSGGYGSGNQGPGRASAAAAAAAIGPGGYGPGSQGSSGTGAYGPGSQGSSGPL<br>SSGPGMSGGYGPGNQGPGGASAAAAAAAASGPGGYGPGSQGPSGPGAYGPGSQGSSGPLSYGTD<br>MSGGYGPGNQGPGGASAAAAAAAASGPGGYGPGSQGSSGPGAYGTGSQGSTGPGSSGPGMSGGY<br>GPGNQGPGGANAAAAAAASGPGGYGPGSQGSSGPGAYGPGSQGSSRPLSSGPGMSGGYEPGNQ<br>GPGGASAAAAAAAASGPGGYGPGSQGPSGPGAYGPGSQGSSGPRSSGPGMSGGYGPGNQGPGGA<br>SAAAAAAASGPGAYGPGSQGSSGPGAYGPGSQGSSGPVSSGPDMSGGYGPGNQGPGGASAAAA<br>AAASGPGGYGPGSQGPSGPGAYGPGSQGSSGPGSSGPGMSGGYEPGNQGPGGASAAAAAAASG<br>PGGYGPGSQGPSGPGAYGPGSQGSSGPGSSGPGMSGGYGPGNQSTGGASAAAAAAASXPGAYG<br>PGSQGSSGPGAYGPGSQGSSGPGSYGPGMSGGYGPGNQGPGGASAAAAAAAASGPAGYGLGSQG<br>SSGPGSSGPGMSGGYGPGNQGLGRASAVAAAAASGPGGYGPGSQGPSRPVAYGPGSQGPSVPG<br>AYGPGSQGSSGPGSSGPGVSGSYGPGNQGPGGASAAAAAAASGPGAYGPGSQGSSGPGSFGPG<br>MSGGYGPGNQGPGGASAAADAAASGPGGYGPGSQGPSGLGAYGPGSQGSSGPGSSGPGMSGGY<br>GPGNQGPGGASAAAAASGPGSQGPSVLGGYGPGSQGPSGWGSQGPSAPSGYGPSASVSASAAA<br>SRLSSPAASSRVSSAVSSLVSSGPTSGAAVSGALNGLVSQISSNNPGLSGCDVLVQALLELVS<br>ALVAILGSASIGAVDYNSVGQTTQTISQYFS | UZU68251.1 |
| **BLAST2** | 1.3 | MSYPRLVLAFLALLSTHALFAAAGGQTPWDTPTLADNFMKCFMNEIGNSGAFTSNQVDDMSTI<br>GDTMMDSVNRLASSGRISKSKLQALNMAFASSMAEIAATEEGGLSIGSKTNAIASALRGAFLQ<br>TTGYSNEQFINEITSLVSMIAEANVNTVSASASAYAGGGYGGSSYGSSSVNSASAAATGPAQQ<br>GPGSYGPSSVPGGYGPSGSSAAAAASGGQGLGNYGPSGSGGAGPSGTGGYGPGSQGPSRPSGP<br>GAAAAAAAASGPGGYGPSAPSGPGSQGPSGPSGSGASAAAAAASGSGGYGSGSQGPSGPGSQG<br>PSGPGTSAAAAAAANGPGGYGPGSQGPSGPGGYRPGSQGPSGPGSSGPGMSGGYGPGNQGPGG | UZU68251.1 |
| | | ASVAAAAAGSGPGGYGPGSQGPSGPGSSGPGMSGGYGPGNQGPGGASAAAAAASGPGGYGPGS<br>QGSSGPGAYGPGSQGSSGPGSSGPGMSGGYGPGNQGPGGASAAAAAAASGPGGYGPXSKGPGR<br>HXGPGSSGPGMSGGYGPGNQGPGGASAAAAAAASGPGGYGPGSQGPSGPGAYGPGSQGSSGPG<br>SSGPGMSGGYGPGNQGPGGASAAAAAAASGPGGYGPGSQGPSGPGAYGPGSQGSSGPLSSGPD<br>MSGGYGPGNQGPGGASAAAAAAASGPGGYGPGSQGPSGPGAYGPGSQGSSGPGSSGPGMSGGY<br>GPGNQGPGGASAAAAAVASGPGGYGPGSQGSSGTGAYGTGSQGSTGPGSSGPGMSGGYGPGNQ<br>GPGGASAAAAAAASGPGGYGPGSQGSSGPGAYGPGSQGSSRPLSSGPGMSGGYGPGNQGPGGA<br>SAAAAAAASGPGGYGPGSQGPSGPGAYGSGSQGSSGPGSSGPGMSGGYGPGNQGPGRASAAAA<br>AAASGPGGYGPGSQGPSGPGAYGPGSQGSSGPLSYGTDMSGGYGPGNQGPGGASAAAAAAASG<br>PGGYGPGSQGPSGPGAYGPGSQGSSGPGSSGPGMSGGYGPGNQGPGGASAAAAAEASGPGAYG<br>PGSQGSSRLLSSGPGMSGGYGPGNQGPGGASAAAYAAASGLGGYGPGSQGPSGPGAYGSGSQG<br>SSGPGSSGPGMSGGYGSGNQGPGRASAAAAAAAIGPGGYGPGSQGSSGTGAYGPGSQGSSGPL<br>SSGPGMSGGYGPGNQGPGGASAAAAAAASGPGGYGPGSQGPSGPGAYGPGSQGSSGPLSYGTD<br>MSGGYGPGNQGPGGASAAAAAAASGPGGYGPGSQGSSGPGAYGTGSQGSTGPGSSGPGMSGGY<br>GPGNQGPGGANAAAAAAASGPGGYGPGSQGSSGPGAYGPGSQGSSRPLSSGPGMSGGYEPGNQ<br>GPGGASAAAAAAASGPGGYGPGSQGPSGPGAYGPGSQGSSGPRSSGPGMSGGYGPGNQGPGGA<br>SAAAAAAASGPGAYGPGSQGSSGPGAYGPGSQGSSGPVSSGPDMSGGYGPGNQGPGGASAAAA<br>AAASGPGGYGPGSQGPSGPGAYGPGSQGSSGPGSSGPGMSGGYEPGNQGPGGASAAAAAAASG<br>PGGYGPGSQGPSGPGAYGPGSQGSSGPGSSGPGMSGGYGPGNQSTGGASAAAAAAASXPGAYG<br>PGSQGSSGPGAYGPGSQGSSGPGSYGPGMSGGYGPGNQGPGGASAAAAAAASGPAGYGLGSQG<br>SSGPGSSGPGMSGGYGPGNQGLGRASAVAAAAASGPGGYGPGSQGPSRPVAYGPGSQGPSVPG<br>AYGPGSQGSSGPGSSGPGVSGSYGPGNQGPGGASAAAAAAASGPGAYGPGSQGSSGPGSFGPG<br>MSGGYGPGNQGPGGASAAADAAASGPGGYGPGSQGPSGLGAYGPGSQGSSGPGSSGPGMSGGY<br>GPGNQGPGGASAAAAASGPGSQGPSVLGGYGPGSQGPSGWGSQGPSAPSGYGPSASVSASAAA<br>SRLSSPAASSRVSSAVSSLVSSGPTSGAAVSGALNGLVSQISSNNPGLSGCDVLVQALLELVS<br>ALVAILGSASIGAVDYNSVGQTTQTISQYFS | |
| | 2.3 | LGGQGAGRGAGAAAAAAGGAGQGGYGGLGGQGAGRGAGAAAAAAGGAGQGGYGGLGGQGAGQG<br>AGAAAAAAGGAGQGGYGGLGGQGAGQGAGAAAAAAGGAGQGGYGGLGSQGAGRGGYGGQGAGA<br>AAAAAAGGAGQGGQGLGGQGAAAAAGGAGQGGYGGLGGQGAGAAAAAAGGAGQGGYGGLGGQG<br>AGRGAGAAAAAAGGAGQGGYGGLGGQGAGQGAGAAAAAAGAGGAGQGGYGGLGSQGAGRGGYG<br>GQGAGAAAAAAAAGGAGQGGQGLGGQGAAAAAGQGGYGGLGGQDAGQGAGAAAAAAGGAGQGGY<br>GGRGGAGRGGYGQGAGAAAAAAAGGAGQGGYGGLGGQGAGQGAGAAAAAAGGAGQGGYGGLGGQ<br>GAGAAAAAAAGGAGQGGYGGLGGQGAGQGAGAAAAAAGGAGQGGYGGLGSQGAGRGGYGGQGAG<br>AAAAAAAGGAGQGGQGLGGQGAAAAAGDAGQGGYGGLGGQGAGAAAAAAGGAGQEGYGGLGGQ<br>GAGRGAGAAAAAAGDAGQGGYGGLGGQGAGQGAGAAAAAAGGAGQGGYGGLGGQGAGQGAGAA<br>AAAAGGAGQGGYGGLGSQGAGRGGYGGQGAGAAAAAAAAGGAGQGGQGLGGQVAAAAAGGAGQG<br>GYGGLGGQGAGAAAAAAGGAGQGGYGGLGGQGAGRGAGAAAAAAGGAGQGGYGGLGGQGAGQG<br>AGAAAAAAGGAGQGGYGGLGGQGGLGGQGAGQGAGAAAAAAGGAGQGGYGGLGGQGAGQGAGA<br>AAAAAGGAGQGGYGGLGSQGAGRGGYGGQGAGAAAAAAAAGGAGQGGQGLGGQGAAAAAGGAGQ<br>GGYGGLGGQGAGAAAAAAGGAGQGGYGGLGGQGAGRGAGAAAAAAGGAGQGGYGGLGGQGAGR<br>GGGGGAAAAAAAGGAGQGGYGGLGGQGAGQGAGAAAAAAGGAGQGGYGGLGTQGAGRGGYGGQG<br>AGAAAAAAAAGGAGQGGQGLGGQGAAAAAGGAGQGGYGGLGGQGAAAAAAAAGGAGQGGYGGLGG<br>QGDGQGAGAAAAAAGGAGQGGYGGLGSQGAGRGGYGGQGAGAAAAAAAAGGAGQGGQGLGGQGA<br>AAAAGGAGQGGYGGLGGQGAGAAAAAAGAAAAAGGAGQGGYGGLGGQGAGQGAGAAAAAAGGA<br>GQGGYGGLGGQGAGQGAGAAAAAAGGAGQGGYGGLGSQGAGRGGYGGQGAGAAAAAAAAGGAGQ<br>GGQGLGGQGAAAAAGGAGQGGYGGLGGQGAGAAAAAAGGAGQGGYGGLGGQGAGRGAGAAAAA<br>AGGAGQGGYGGLGGQGAGQGAGAAAAAAGGAGQGGYGGLGGQGAGQLAGAAAAAAGGAGQGGY<br>GGLGSQGAGRGGYGGQGAEAAAAAAAAGGAGQGGQGLGGQGAAAAAAGGAGQGGYGGLGGQGAG<br>AAAAAAGGAGQGGYGGLGSQGAGRGAGAAAAAAGGAGQGGYGGLGGQGAGRGAGAAAAAAAGGAA<br>QGGYGGLGSQGAGAAAAAAGSAAGGLGGYGGLGGQGAGQGGYGGVGSGASAASSAASRLSSPE<br>ASSRVSSAVSNLVSSGPTNSAALSSTISNVVSQIGASNPGLSGCDVLVQALLEVVSALIHILG<br>SSSIGQVNYGSAGQATQIVGQSIYQALG | GFY35630.1 from[5] |
| | 3.3 | MNWSIRLALFGFVVLSTQTVFAVGQAATPWENSQLAEDFINSFLRFIAQSGAFSPNQLDDMSS<br>IGDTLKTAIEKMAQSRKSSKSKLQALNMAFASSMAEIAVAEQGGLSLEAKTNAIASALASAFL<br>ETTGVVNQQFVSEIKGLIYMIAQASSNEISGSASGSGGGSGGGGGGGGYGPGSYASASVAAA<br>YGSAPQGAGGPSPQGPSQQAPISQGPYGPGAAAAAAASGGYGPGAGQQGPSGGGQQGPGGAGQ<br>QGPGGQGPYVPSAAAAAAGGYGPGAGQ | AWK58646.1 from[3] |
| | 4.3 | MSCPRLVLAFLALLSTNALFAAAAAATPWDSPALADSFMKSFMDGIGTSGAFTSSQIDDMSTI<br>GDTMMDSVNRLASSGRISKSKLQALNMAFASSMAEIAATEEGGLSIGAKTSAIASALRGAFLQ<br>TTGYANEQFINEITSLINMIAQANVNAVSASASASAGGGYGAPAYGPSSYGPSQQQSSASSVS<br>VSASAAGAGPRSQAPSRPAQQGPRGYGPSGPGGTAAASASAGGPGSQGPYGPGQQGPGPRGPS<br>RPSQQGPGGYGPSGPGGASAAAAAAAAGGPGGQGPYGPGQQGPGAGPYGPGQQGPGQQGPGGY<br>GPSGPGGAAAAAAAAAAAGGPGGQGPSGPGQQGPGGYGPSGPSGASAAAAAAGGQGPYGQGQQG<br>PRGYGPSGPGGTAAAAAAAAGGPGGQGQYGPGQQGPGGYGSSGTGGASAAAAAAAAGGPGGQGP<br>YGPGQQGPYGPGQQGPGGQGRGGYGPSGPGGASAAAAAAAAGGPGGQGQYGPGQQGPGGYGPS<br>GPGGASAAAAAASAAGGPGGQGPSGPGGQGPSGPGQQGPGGYGPSGPSGASAAAAVAGGQGPYG<br>QGQQGPGGYGPSGPAGASAASAAAAAGGQGGQGPYGPGQQGPYGPGQQGPGGQGRGGYGPSGP<br>GGASAAAAAAAAAGGPGGRGEYGPGQQGPGGYGPSGPGGASAAAAASAAGGPGGQGPSGPGQQG<br>PGGYGPSGPSGASAAAAVAGGQGPYGQGQQGPGGYGPSGPAGASAASAAAAAGGQGGQGPYGP<br>GQQRPYGPGQQGPGGQGRGGYGPSGPGGASAAAAAAAAGGPGGQGQYGPGQQGPGGYGPSGPG<br>GASAAAAASAAGGPGGQGPSGPGQQGPGGYGPSGPSGASAAAAVAGGQGPYGQGQQGPGGYGP<br>SGPAGASAASAAAAAGGQGGQGPYGTGQQGPYGPGQQGPGGQGRGGYGPSGPGDASAAAAAAA<br>AGGPGGQGQYGPGQQGPGGYGPSGPGGASAAAAASAAGGPGGQGPSGPGQQGPGGYGPSGPSG<br>ASAAAAVAGGQGPYGQGQQGPGGYGPSGPAGASAASAAAAAGGQGGQGPYGPGQQGPYGPGQQ<br>GPGGQGPGGYGPSGPGGASAAAAAAAAGGPGGQGQYGPGQQGPGGYGPSGPGGASAAAAASAA<br>GGPGGQGPSGPGQQGPGGYGPSGPSGASAAAAVAVGQGPYGQGQQGPGGYGPSGPAGASAASA<br>AAAAGGQGGQGPYGPGQQGPYGPGQQGPGQQGPGGYGPSGPGGSAAAAAAAAAGGPGGQGPSG<br>PGQQGPGGYGPSGPSGASAAAAAAAGGQGPYGQGQQGPGGYGPSGPAGASAASAAAAAGGQGGQ<br>GPYGPGQQGPYGPGQQGPGGQGPGGYGPSGPGGASAAAAAAAAGGSGGQGPYGPGQQGPYGPG<br>QQGPGQQGPGGYGPSGPSGASAAAAAAAGGQGPYGQGQQGPGGYGPSGPAGASAASAAAAAGGQ<br>GGQGPYGPGQQGPYGPGQQGPGGQGPGGYGPSGPGGASAAAAAAAAGGSGGQGPYGPGQQGPY<br>GPGQQGPGQQGPGGYGPSGPSGASAAAAAAAGGQGPYGQGQQGPGGYGPSGPAGASAASAAAAA<br>GGQGGQGPYGPGQQGPYGPGQQGPGGQGPGGYGPSGPGGASAAAAAAAAGGSGGQGPYGPGQQ<br>GPYGPGQQGPGQQGPGGYGPSGPGGTAAAAAAAAAAGGPAGQGPSGPGQQGPGGYGPSGPSGAS<br>AAAAAAGGQGPYGQGQQGPGGYGPSGPAGASAASAAAAAGGQGGQGPYGPGQQGPYGPGQQGP<br>GGQGPGGYGPSGPGGASAAAAAAAAGGSGGQGPYGPGQQGPYGPGQQGPGQQGPGGYGPSGPS<br>GASAAAAVAGGQGPYGQGQQGPGGYGPSGPAGASAASAAAAAGGQGGQGPYGPGQQGPYGPGQ<br>QGPGGQGPGGYGPSGPGGASAAAAAAAAGGSGGQGPYGPGQQGPGQQGPGGYGPSGPSGASAA<br>AAAAGGQGPYGQGQQGPGGYGSSGSGGAAAAAATAGGPGGQGQYGPGQQGPGGYGPSGPGAAS<br>AAAAAAAAGGPGGQGLSGPGQQGPGGYGPSGPSGASAAAAVAGGQGPYGQGQQGPGGYGPSGP<br>AGVSAASAAAAAGGQGGQGPYGPGQQGPYGPGQQGPGGQGPGGYGPSGPGGASAAAAAAAAGG<br>SGGQGPYGPGQQGPYGPGQQGPGGQGPGGYGPSGPGGSAAAAAAAASAAAGGAGGQGPSGPGQ<br>QGPESYGPSGPSGASAATAAAGGQGPYGQGQQGPGGYGPSGPVSGVSVSVSSAASRLSSPAAS<br>SRVSSAVSTLASSGPSDAGVVSSALSNLVSQVSTNHPGLSECDVIVQALLELVSALVHILGSS<br>SVGQVDYNGASYSAQNLGQAVAQALA | UZZ64708.1 |
| | 5.3 | MSWSTLALAILVVFSTQCIFIAGQANTPWSDTATADAFIQNFLGAVSGSEAFTPDQLDDMATV<br>GDTIMSAIDKMARNNKSSKSKLQSLKMAFASSIAEIAAVEQGGQSMDIKTNAIANALDSAFYM<br>TTGSTNQQFVNEMRRLINMLSAASVNELSYGGGTSAAAATAGGYGQGASGYDPGSSPASAAAL<br>KGPSKREPSGLGAAAAAPSEYGSSQQGPSGTKAATIAAAKRGPTSYGPRQQRPGGSGAPAATA<br>GRGPGSYGPEQQGPRGSGAAADEAGPGQQEPGADAAAALGSGSGDQGPGRFDAAAATAKPRGN<br>GPGQQGSGVASAAAAGSGPRGYGPGQQAHRGHGAAAAATGSGGYEPGQQGPGGPGAAAAGLGP<br>GGYGPGKQGQGRPAATAAAAEPGGYGPRIQGTGAAAAAATGRGPGGYGPGQQGPGGSGAVKAA<br>DGPGSFGPGQPGGPGAAATAAARRGPGGYGPEQEPGRPSVAAASAGPGGYGPRQQGPGGYAPG<br>QQEPGVPGATGAAAAGRGSGYANGKKVPGGPGAAAAAATGSTPGAYGPGQQGPGGDDPKQQAP<br>ASSSATEAAAGPRGYGPGKQGPGAAVAAAAGSGPGGYGPRQQGPGGPGIGPGVYGPGQQGKRG<br>YGPGQQGPGGFGAAAAAAAAAGPGGYGPGQEGPGSAVAAAAGREPGGYGPKQKGAGATAAAAA<br>ESKPGGDGPEQQRPGGPGAAAPAKPGRGPGSYGPGQQGPRSSGVAAATAAGPGDYGPDKRGPG<br>GPGVAAAGRGPGRPGSAADANAGSPGGYGPGQQGPGGSGSGPGVYRPRQSGGPGAAVAAATRR<br>GYGYGQGQQGPEGPGAVTAAAAGSGPGGYGTGQQGKEGYGSREQEPGDSGSAAAAFGPGVSGP<br>KQQGPGEKAAPASGSGTRGDGPGQQGPGGSGAAVAAEAGRGSGGYEPGQQGPEGSGAAAAAAS<br>RPGGYGLGQEGPGSAASTAAGRGIGGHGPGQQGPGGPGAAAAAATGRGPGGYNPGQKGPAVYG<br>TKQGPEEPRSDAAATTGTGPGQQGPGGPVTAAVAAGSGQQKLSAAAAAAAGRGPGGYGPGQQG<br>PAATATAAGRGLGGYGPGQQGPRGTGAAAEAAAGRGPGGYGPGQQEAGVSGEAAEAAGPGPPP<br>QGPGTAAIAAAGSVPGGYVPGQRGTGGPAAAAATGLGGYKPGQQGPGGYAPGQKGPEAAAAGR<br>GSGYGPGKQVPGGPGAAAAAAEPGPPGEYGTEKRGPKGDGPKQQAPAGSSAAAAATGPQGYGP<br>GQQGPGATASAAAGSRPVRYGPGQKGPGAGPKGYEPGQQGPGGPGSAAAAAEPGGYGPAQQGP<br>GVPGAAAGRRGLGYGPGKHGPSAAAAAAAGSGPGGYGPGQQGKGGYGPGKQEPGNFGAAAAAS<br>GPGGYRPGKEGPGSAVAAAARRGPGGYGPKQKGAGAMAAEGPGVSGAAAATTSGPGGYGAGQE<br>GPGAVAVATAGRGPGGYRPGLYGPGGYGSAAEAAGPGGYGSKQQGTISTAAAAGSEPGRYGP<br>GQQGPGGSGVAAAAEERREPGDYRPGQQGPGGPSVAAASAGLGGYGPGNKDQEDLVAAASAGL<br>GGYGPRQQGPGGYAPGQQGPGGYAPGRQGPGVPGAAAAAGAGSGYRPGQQVPGGPGTTAAAAA<br>GSTSGEYGPGQQGPKVDGPKQQAPAGSSDAAAAAGPRGYGPGQQGPVAAASAAAGSGPVGYGP<br>GQRGPGAAVAAAAGSGPGGYGPRQQGQVGHGRAATAEAGRGPEVYEPGEQGPGRLGSAAAAAG<br>PGGYRPRQQGPGVHGAATARRGSGYGPAQQGPEAPGAAAAAAAAGSGPGGYGPGKQGKGGYGPG<br>QQGPGDFGAAAAASGPGGYGPGSAAAAAAGRGPGGYGPKQQGAGAMASTAAGSIPGGYGPGQQ<br>GPGDFGAAAAAAASGPGGYGPGQEGPGSAAAAAAGRGPGGYRSGQQEPGGFGSTAAAAGPGGYG<br>PGQQGPGTVAVAAAGSGPGGYGTGQQGPGGSSAAAASAGPGGYGPGQQGPGVPGAAASAAAVR<br>GSGYGARQQVPGGPGAAAAAVTGRRPGGYGPGQQGPGRLDAASAAVGPGSYGPEQQGPGAAAA<br>GRGPGDYGTEQQGPGRYGTGQQGPGRPVTAAVDSGSEQQGLSAAAAAAAGRGNGGYLPGQQGP<br>AVAAAAAGRGRGGYGRQGPGGPGAALANAGPEGYGPGQQGTDAAAATAAVSGPGAASSTGRSP<br>EGYGSEQQGPAGPGAATAAAAGRVPGGYRSGEQGPEGPGAAAATVAGIGPGGYGSRQEGPGGP<br>VAAAAASGPGGYRPGQPGGPVATAAAAGRGPRGYVPGQQGPVGAAAATSRSGPGSGPGKQGPG<br>AAAAAAGPGGYGPEEQGPGAALAAAAGSGPGGYGPGPQASAAVSRLAFPDSRSRVSSASSNLV<br>ASGPTNSAALSNAISNTVSEIGASYPGLSGCDVLVQALMEIVSALVAILSSSSIGQVNYVAVS<br>QSAQVVSQSLLQALY | GFY34551.1 from[5] |

**Table S2. Assessment of novelty and protein type for generated sequences**, by comparing similar existing protein sequences from BLAST searches. For the evaluation of novelty, the BLAST search results are arranged in descending order based on query cover (QC). Details encompassing the highest values and the common value ranges for both query cover (QC) and percent identity (id%) are presented. QC serves to assess the alignment of sequences, while id% signifies the degree of composition similarity. Sequences exhibiting similarity values below 50% to 60% are considered novel according to established criteria[6,7]. In summary, the generated sequences are evaluated to be novel within the broader protein dataset, with analysis provided in the table, and are classified as MaSp through similar sequence comparison.

| Set | Highest value of [QC, id%] | Common value range of [QC, id%] | Novelty Analysis |
|---|---|---|---|
| 1 | [97%, 38%] | [5%~39%, 38%~62%] | Only one sequence exhibits a high alignment and QC values drop significantly for others. Besides, Overall, there is a notable deficiency composition similarity across all sequences |
| 2 | [11%, 72%] | [8%~11%, 57%~81%] | In general, the sequences display limited alignment, yet reasonable composition similarities are evident |
| 3 | [98%, 46%] | [88%~98%, 38%~62%] | Significant sequence alignments are apparent for the majority of sequences. However, due to the considerably shorter length of the query compared to the searched sequence, the comparison consisten is diminished, leading to lower coherence. Additionally, there is a lac of substantial composition similarity |
| 4 | [18%, 44.5%] | [8%~18%, 44%~87%] | In general, there is a lack of substantial sequence alignments, althou there is a reasonable level of similarity in sequence composition |
| 5 | [98%, 44%] | [10%~55%, 38%~98%] | There is only one instance characterized by a high sequence alignmer while the remaining cases exhibit significantly lower QC values. Furthermore, there is a lack of notable similarities in sequence composition across all instances |

**Table S3. Comparison of generated novel sequences and corresponding existing sequences from BLAST search, for the five sets.** In this table, the sequence length, and protein sequence properties, including molecular weight (MW), instability index (II)[8], and Isoelectric point (pI), obtained through ProtParam[9] (implemented in a Python tool package[10]), are analyzed. Moreover, the local confidence measures (pLDDT) of folding predictions using AlphaFold2[11] are provided.

| Set | ID | source | Len. | GT/PDT | R2 | MAE | MW | II | pI | pLDDT (AF2) |
|---|---|---|---|---|---|---|---|---|---|---|
| 1 | 1.1 | Generated | 1096 | [0.250, 0.252, 0.200, 0.151, 0.310, 0.203, 0.293, 0.265]<br>[0.278, 0.274, 0.182, 0.156, 0.299, 0.215, 0.299, 0.25] | 0.8899 | 0.0146 | 91081.42 | 31.42 | 8.20 | 41.12 |
| | 1.2 | Blast 1 | 1713 | N/A | N/A | N/A | 165101.21 | 59.08 | 6.95 | 44.71 |
| | 1.3 | Blast 2 | 1984 | N/A | N/A | N/A | 170811.62 | 28.55 | 5.74 | N/A |
| 2 | 2.1 | Generated | 926 | [0.550, 0.252, 0.750, 0.151, 0.310, 0.203, 0.293, 0.265]<br>[0.745, 0.141, 0.511, 0.257, 0.314, 0.252, 0.247, 0.257] | 0.5640 | 0.0947 | 77765.79 | 32.81 | 8.27 | 41.78 |
| | 2.2 | Blast 1 | 1713 | N/A | N/A | N/A | 159478.20 | 47.48 | 9.87 | 40.93 |
| | 2.3 | Blast 2 | 1603 | N/A | N/A | N/A | 125416.40 | 13.67 | 9.41 | N/A |
| 3 | 3.1 | Generated | 312 | [1.000, 0.252, 0.450, 0.151, 0.200, 0.203, 0.293, 0.265]<br>[0.696, 0.356, 0.363, 0.215, 0.247, 0.152, 0.457, 0.192] | 0.7167 | 0.1118 | 29754.17 | 37.68 | 4.88 | 59.40 |
| | 3.2 | Blast 1 | 367 | N/A | N/A | N/A | 35845.68 | 32.74 | 5.77 | 53.80 |
| | 3.3 | Blast 2 | 279 | N/A | N/A | N/A | 26864.21 | 44.49 | 4.88 | N/A |
| 4 | 4.1 | Generated | 869 | [0.900, 0.252, 0.800, 0.151, 0.310, 0.203, 0.293, 0.265]<br>[0.745, 0.141, 0.511, 0.177, 0.316, 0.243, 0.316, 0.247] | 0.7843 | 0.0835 | 69899.23 | 22.41 | 7.84 | 46.62 |
| | 4.2 | Blast 1 | 648 | N/A | N/A | N/A | 52886.51 | 22.50 | 7.98 | 40.37 |
| | 4.3 | Blast 2 | 2357 | N/A | N/A | N/A | 202052.69 | 29.90 | 7.91 | N/A |
| 5 | 5.1 | Generated | 521 | [0.240, 0.252, 0.206, 0.151, 0.270, 0.203, 0.293, 0.265]<br>[0.278, 0.274, 0.182, 0.157, 0.263, 0.215, 0.316, 0.272] | 0.7751 | 0.0174 | 43816.17 | 28.23 | 6.10 | 46.48 |
| | 5.2 | Blast 1 | 1984 | N/A | N/A | N/A | 170811.62 | 28.55 | 5.74 | 38.40 |
| | 5.3 | Blast 2 | 2472 | N/A | N/A | N/A | 224113.32 | 29.07 | 9.55 | N/A |

**Table S4. Summary of significant motifs.** The motif IDs are identified and used in this work. These motifs are selected and concluded from Table 1 in[1]. Here we assume the specific MaSp type of sequences has minimal impact on the motif analysis.

| ID | Motif | Property | Effect |
|---|---|---|---|
| T_pos_1 | GYGQGG | Toughness | +ve |
| T_pos_2 | GGGQ | | +ve |
| T_neg_1 | SQGP | | -ve |
| T_neg_2 | SY | | -ve |
| T_neg_3 | SV | | -ve |
| TS_pos_1 | GYGQGG | Tensile strength | +ve |
| TS_pos_2 | QGGS | | +ve |
| TS_neg_1 | PQ | | -ve |
| SB_pos_1 | GYGQGG | Strain at break | +ve |
| SB_pos_2 | QGP | | +ve |
| SB_pos_3 | PGA | | +ve |
| E_pos_1 | PA | Elastic modulus | +ve |
| E_pos_2 | GQ | | +ve |
| E_neg_1 | GGQ | | -ve |

**Table S5. Normalization scaling parameters along with units.** The minimum and maximum values of the eight properties, which are used for normalization, are summarized here.

| **Property** | **Unit** | **Min.** | **Max.** |
|---|---|---|---|
| toughness | $GJ/m^3$ | 0.005 | 0.39 |
| SD of toughness | $GJ/m^3$ | 0.001 | 0.136 |
| E | GPa | 0.38 | 37.0 |
| SD of E | GPa | 0.03 | 9.76 |
| strength | GPa | 0.17 | 3.33 |
| SD of strength | GPa | 0.01 | 0.8 |
| strain | % | 5.1 | 53.2 |
| SD of strain | % | 0.1 | 13.7 |

**Supplementary References**

1. Arakawa, K. *et al.* 1000 spider silkomes: Linking sequences to silk physical properties. *Sci Adv* **8**, eabo6043 (2023).

2. Kono, N. *et al.* Darwin's bark spider shares a spidroin repertoire with Caerostris extrusa but achieves extraordinary silk toughness through gene expression. *Open Biol* **11**, 210242 (2021).

3. Collin, M. A., Clarke, T. H., Ayoub, N. A. & Hayashi, C. Y. Genomic perspectives of spider silk genes through target capture sequencing: Conservation of stabilization mechanisms and homology-based structural models of spidroin terminal regions. *Int J Biol Macromol* **113**, 829–840 (2018).

4. Gatesy, J., Hayashi, C., Motriuk, D., Woods, J. & Lewis, R. Extreme Diversity, Conservation, and Convergence of Spider Silk Fibroin Sequences. *Science (1979)* **291**, 2603–2605 (2001).

5. Kono, N. *et al.* Multicomponent nature underlies the extraordinary mechanical properties of spider dragline silk. *Proceedings of the National Academy of Sciences* **118**, e2107065118 (2021).

6. Ni, B., Kaplan, D. L. & Buehler, M. J. Generative design of de novo proteins based on secondary-structure constraints using an attention-based diffusion model. *Chem* **9**, 1828–1849 (2023).

7. Quan, L., Wu, T. & Lyu, Q. Computational de novo protein design: From secondary to primary, then toward tertiary structures. *Chem* **9**, 1625–1627 (2023).

8. Guruprasad, K., Reddy, B. V. B. & Pandit, M. W. Correlation between stability of a protein and its dipeptide composition: a novel approach for predicting in vivo stability of a protein from its primary sequence. *Protein Engineering, Design and Selection* **4**, 155–161 (1990).

9. Gasteiger, E. *et al.* Protein Identification and Analysis Tools on the Expasy Server; in *The Proteomics Protocols Handbook, Humana Press* 571–607 (2005).

10. Official git repository for Biopython: ProtParam.py. *Github*.

11. Jumper, J. *et al.* Highly accurate protein structure prediction with AlphaFold. *Nature* **596**, 583–589 (2021).